\documentclass[twocolumn,aps,prc,amsmath,showpacs,superscriptaddress,nofootinbib]{revtex4}
\usepackage{graphicx,epsfig,longtable}
\usepackage{color}

\newcommand{\beqy}{\begin{eqnarray}}
\newcommand{\eeqy}{\end{eqnarray}}
\newcommand{\bmlet}{\begin{subequations}}
\newcommand{\emlet}{\end{subequations}}
\newcounter{saveeqn}

\def\ga{\,\,\raise0.14em\hbox{$>$}\kern-0.76em\lower0.28em\hbox  
{$\sim$}\,\,}  
\def\la{\,\,\raise0.14em\hbox{$<$}\kern-0.76em\lower0.28em\hbox  
{$\sim$}\,\,}

\begin{document}
%\begin{frontmatter}

\title{Improved microscopic nuclear level densities within the triaxial Hartree-Fock-Bogoliubov plus combinatorial method}

\author{S. Goriely}
\email{sgoriely@astro.ulb.ac.be}
\affiliation{Institut d'Astronomie et d'Astrophysique, ULB - CP226,  1050 Brussels, Belgium}
\author{W. Ryssens}
\affiliation{Institut d'Astronomie et d'Astrophysique, ULB - CP226,  1050 Brussels, Belgium}
\author{S. Hilaire}
\affiliation{CEA/DIF, Service de Physique Nucl\'eaire, BP 12, 91680 Bruy\`eres-le-Ch\^atel, France}
\author{A.J. Koning}
\affiliation{Nuclear Data Section, IAEA, Wagrammerstrasse 5, 1400, Vienna, Austria}

\date{\today}

\begin{abstract}
 New developments have been brought to our energy-, spin- and parity-dependent nuclear level 
 densities based on the microscopic combinatorial method. 
Our new calculation is based on the BSkG3 mean-field model \cite{Grams23} which relies on a three-dimensional coordinate-space representation of the nucleus, allowing for the spontaneous breaking of ground state rotational, axial and reflection symmetry.
In particular, we now account for the impact of possible triaxial deformation of nuclear ground states on the level density.
This has two effects on our calculations:  the additional freedom of the single-particle levels  affects the intrinsic level density
while the absence of a rotational symmetry axis results in a larger collective correction.
% The intrinsic nuclear level density is found to be affected mainly by the different single-particle level density at the
% triaxial deformation as well as the additional contribution stemming from the collective effects.
The present model reproduces the experimental s- and p-wave neutron resonance spacings with a degree of accuracy comparable to that of the best global models available. It is also shown that the model gives a reliable extrapolation at low energies where experimental data
on the cumulative number of levels can be extracted. The predictions are also in good agreement with the experimental data extracted from the Oslo method. 
Total level densities for more than 8500 nuclei are made available in a table 
format for practical applications. For the nuclei for which experimental s-wave spacings and enough low-lying states exist, renormalization factors are provided to reproduce simultaneously both observables. 
The same combinatorial method is used to estimate the nuclear level densities at the fission saddle points of actinides and at the shape isomer deformation. Finally, the new nuclear level densities are applied to the calculation of radiative neutron capture cross sections and compared with those obtained with our previous combinatorial model.
\end{abstract}

%\begin{keyword}
%NUCLEAR STRUCTURE: nuclear level density, nuclear energy levels, Hartree-Fock
% PACS Numbers: 21.10.Ma (level density), 
%               21.10.-k (Properties of nuclei; nuclear energy levels), 
%               21.60.Jz (HF & QRPA)
%\end{keyword}

%\end{frontmatter}
\maketitle

\section{Introduction}
\label{sec:intro}

 The study of nuclear level densities (NLDs) goes back to 1936 with Bethe's pioneering work \cite{Bethe36}.
% Since then, several improvements of varying sophistication have been developed:
 %to reproduce the available experimental data.
The so-called partition function method is by far the most widely used technique,
 %to calculate level densities,
 probably because of its capacity to provide simple, though non-predictive, analytical formulae. This method corresponds to the zeroth order approximation of a Fermi gas model. 
%and thus \w{results in simple analytical}, though non-predictive, expressions for the NLD.
 %ability to provide simple analytical formulae.
 %Such a model corresponds
 %of a Fermi gas model and leads to simple analytical, though non-predictive, expressions for the NLD.
Over the years,  various phenomenological modifications to the original analytical
 formulation of Bethe have been suggested to match certain features of the experimental data, in particular by including shell, pairing and collective effects~\cite{Koning08,Goriely96,Capote09}.
 However, drastic approximations are usually made in deriving such analytical
 NLD formulae and their shortcomings in matching experimental data are overcome by empirical parameter adjustments. 
% Despite a steady increase over the past few decades, the relative paucity of experimental information - particularly at high excitation energy - remains a major challenge faced by NLD models and the associated parameter fits.
More specifically, such empirical systematics based on simple analytical formulae adjusted to the neutron resonance spacings at the neutron separation energy of (nearly-)stable nuclei
necessarily lead to biased NLD predictions, particularly when extrapolating to high excitation energies (typically $U \ga 10~{\rm MeV}$),
spins differing significantly from those involved in the experiment and/or to exotic nuclei. 
Barring a dramatic increase in the availability of experimental data, such extrapolations will remain required to produce the significant amount of nuclear data that is required by applications such as nuclear astrophysics and the design of accelerator-driven systems. Ideally, such extrapolations should be based on a nuclear model
that is both {\it reliable} and {\it accurate}.
Even if it is not yet possible to describe NLD from first principles, the extrapolations of
physically sound models that are rooted firmly in a microscopic description of the nucleus have to be preferred to those of more phenomenological highly-parametrized models.

Several different types of microscopic NLD models have been developed over the years, see
e.g. Refs.~\cite{Goriely08b,Hilaire12,Uhrenholt13,Dossing19,Alhassid15,Alhassid15b,Zelevinsky18,Zhao20,Hilaire23}
and references therein. Historically, they have seldom been used for practical applications due to
three seemingly apparent flaws:
({\it i}) their lack of global accuracy w.r.t. experimental data,
({\it ii}) their limited flexibility in comparison with highly parametrized analytical expressions to match the data by tuning parameters,
and ({\it iii}) their associated computational cost that renders large-scale applications difficult.

These flaws are no longer relevant today, as increases in available computing power have rendered
   the global application of two different types of microscopic NLD model possible: the statistical
   approach of Ref.~\cite{Demetriou01} and the combinatorial modelling of Refs.~\cite{Goriely08b,Hilaire12},
   both of which are based on energy density functional (EDF) calculations of nuclear ground states.
   The mean field plus statistical approach of Demetriou et al. \cite{Demetriou01} first established that such a bottom-up modelling of NLD can
   be competitive with more phenomenological models w.r.t.~the reproduction of experimental data and
   is thus suitable for practical applications. Unfortunately, a statistical approach
   is inherently unable to describe the parity dependence of the NLD, nor is it suited to describe the  
   discrete  levels at low energies.

%    The combinatorial approach followed in Refs. \cite{Goriely08b,Hilaire12} clearly demonstrated that these drawbacks can be overcome and that microscopic models can compete with empirical formulas in the global reproduction of experimental data.
% A global microscopic NLD prescription within the statistical approach based on the Hartree-Fock-BCS (HFBCS)
% ground state properties~\cite{Demetriou01} has proven the capacity of microscopic models to compete with
% phenomenological models in the reproduction of experimental data and consequently to be adopted for practical
%  applications. However, this statistical approach presents the drawback of not describing the parity dependence
%  of the NLD, nor the discrete (i.e non-statistical) nature of the excited spectrum at low energies.

Combinatorial modelling is competitive w.r.t. the global description of data, yet it does not suffer from these deficiencies: it provides the energy, spin and parity dependence
of NLD and is more suited to predict the discrete spectrum at low energies~\cite{Goriely08b,Hilaire12}.
This non-statistical regime can have a significant impact on calculated cross sections, particularly for processes
known to be sensitive to spin or parity distributions such as isomeric production or low-energy neutron
capture~\cite{Goko06,Cannarozzo23}. This approach can also provide the partial particle-hole (PH) level density, a quantity that 
is of particular relevance in the description of the pre-equilibrium reaction but that cannot be extracted
in any satisfactory way from the statistical approaches.

The starting point of the combinatorial approach is the nuclear single-particle level (SPL) scheme, obtained from a
mean-field-type calculation of the nuclear ground state. State-of-the-art models today rely on Hartree-Fock-Bogoliubov (HFB)
calculation based on a nuclear EDF such that NLDs can be predicted consistently for all nuclei.
By counting the number of possible PH excitations based on the SPL scheme resulting from a HFB calculation, the combinatorial method constructs the density of \emph{intrinsic} states as a function of excitation energy, angular momentum and parity. By building
collective excitations on top of the intrinsic states, one constructs the physical states. The sheer number of states involved ensures that this second step cannot rely on a full-fledged many-body calculation; instead, one uses simple phenomenological recipes with microscopic input from the ground-state calculation. The classic example is rotational
collectivity: a rigid rotor model allows the explicit inclusion of the rotational excitations of all intrinsic states into the NLD calculation, assuming that the moment of inertia (MOI) of the ground state is representative.

The way a nuclear ground state is modeled thus strongly affects the calculation of NLDs in the combinatorial approach.
This is especially evident when considering symmetry assumptions: a self-consistent symmetry imposed on the
nuclear configuration will (a) strongly affect the spl, usually by allowing the assignment of quantum numbers to single-particle states,
and (b) stop the nucleus from engaging in the collective motion associated with said symmetry. However, any conserved symmetry
significantly simplifies the solution of a mean-field solution. This simplification is usually welcome when attempting
a global description of nuclear ground states: most large-scale microscopic models allow nuclei to break rotational
symmetry but impose axial symmetry as well as time-reversal and parity. Although such a choice can account for the
dominant strain of collectivity across the nuclear chart, it cannot account for the appearance of triaxial deformation and
the associated rotational spectrum nor for the doubling of rotational bands in an octupole deformed nucleus.

We have recently started building a series of large-scale models that do away with this restriction:
the Brussels-Skyrme-on-a-Grid (BSkG) models do not impose axial symmetry on the nucleus and consistently allow
all nuclei the freedom to take on triaxial shapes~\cite{Scamps21}. Starting with the second entry in the series - BSkG2 - we also
systematically included the effects of time-reversal breaking on the ground states of odd-mass and odd-odd nuclei~\cite{Ryssens22}.
For BSkG3, we also included the effect of left-right asymmetry on nuclear ground states~\cite{Grams23}.
Obtaining the NLDs these models predict thus requires some refinement of the combinatorial approach: simultaneously accounting
for time-reversal symmetry breaking as well as octupole and triaxial deformation. As will be explained below, time-reversal breaking
does not significantly affect NLDs while the inclusion of reflection asymmetry is straightforward; including triaxial deformation
requires more work.

The combinatorial formalism is described in Sect.~\ref{sect_comb} where the nuclear structure properties predicted by BSkG3 and the impact of triaxiality on NLD are also discussed in details. In Sect. ~\ref{sect_res}, the resulting NLD are compared with experimental
data. A possible renormalization of the NLD on such data is included. Some applications to reaction cross section calculations are illustrated in Sect.~\ref{sect_xs}. The application of the present model to the calculation of the NLD at the fission saddle points is discussed in Sect.~\ref{sect_fis}.
Conclusions are finally drawn in Sect.~\ref{sect_conc}.

\section{A symmetry-broken combinatorial NLD model}
\label{sect_comb}

\subsection{Nuclear ground states with BSkG3}
\label{sect_triax_gs}

The BSkG models mentioned in the introduction are large-scale models of nuclear structure that aim to describe the largest amount of observables \emph{simultaneously} across the entire nuclear chart starting from a description of the nucleus that is as microscopic as possible~\cite{Scamps21}. The starting point of these models are EDFs of the Skyrme type, whose coupling constants were adjusted to, among other things, essentially all known masses, charge radii, the properties of infinite nuclear matter and even empirical fission barriers of actinide nuclei. Contrary to most global models in the literature, the BSkG models rely on a three-dimensional coordinate space representation of the single-particle wavefunctions such that they allow the nucleus to explore the spontaneous breaking of
 rotational, axial, reflection and time-reversal symmetries.

We focus here on BSkG3, as opposed to the most recent model BSkG4. The latter model improves on the description of the pairing channel for extremely neutron-rich nuclei, but BSkG3 is at the heart of our current efforts to predict the fission properties of several thousand heavy nuclei. Nevertheless, the framework we develop here is entirely applicable to the more recent model.

Quantitatively, BSkG3 offers accurate predictions of ground state properties on the scale of the nuclear chart, with root-mean-square (rms) deviations of 0.63~MeV on 2457 masses \cite{Wang21} and 0.0237 fm on 810 charge radii \cite{Angeli13}. The model description of fission properties is remarkable: it matches 45 empirical primary fission barriers of actinide nuclei~\cite{Capote09} with an rms deviation of 0.33 MeV and describes all 107 known spontaneous fission half-lives within a factor of $1.5 \times 10^3$~\cite{sanchezfernandez2025}; the combined accuracy for both barriers and half-lives is, to the best of our knowledge, unmatched by any other approach available today. Of particular interest for NLD calculations, is the microscopically founded treatment of nucleon pairing in BSkG3 that is in line with the predictions of advanced many-body methods for infinite nuclear matter. Finally, by adopting an extended form of the Skyrme functional, the model reconciles the description of matter at high densities and at saturation density leading to a stiff equation of state of pure neutron matter, compatible with the observational evidence for heavy pulsars with  $M \ga 2.1 M_{\odot}$.

The large-scale calculation of some 8500 nuclei with BSkG3 showed that triaxial deformation is energetically favorable for many nuclei. The regions of the nuclear chart where triaxial deformation is predicted matches that where spectroscopic evidence for such deformation is available~\cite{Scamps21}; in particular, BSkG3 can describe masses~\cite{hukkanen2023,hukkanen2023a,hukkanen2024} and charge radii~\cite{maass2025} and help us interpret the nature of isomers~\cite{stryjczyk2025} in the $Z \approx 44$ region where experimental evidence for triaxiality is most plentiful. Four islands of octupole deformed nuclei were also found (see Fig.~8 of Ref.~\cite{Grams23}); although experimental evidence for octupole deformation is less numerous, these islands are in the vicinity of the so-called octupole magic numbers~\cite{butler1996}. Because both of these deformations imply the development of specific patterns in the excitation spectrum of a nucleus, these deformations must also be accounted for in NLD calculations~\cite{bjornholm1974,younes2004}, as discussed below. Finally, note that, although the HFB model includes the effect of so-called time-odd terms, those effects are not propagated here to the NLD calculations; we follow the standard equal filling approximation. Indeed, early tests showed that time-reversal breaking shifts the energies of the SPL by at most some hundred keV or less; these changes have significant impact on several structural properties of the nucleus such as its magnetic moment, but the energy differences are typically too small to make a meaningful impact on the NLD while the breaking of Kramers two-fold degeneracy complicates the combinatorial calculations.

\begin{figure*}[htbp]
 \begin{center}
 \includegraphics[scale=0.45]{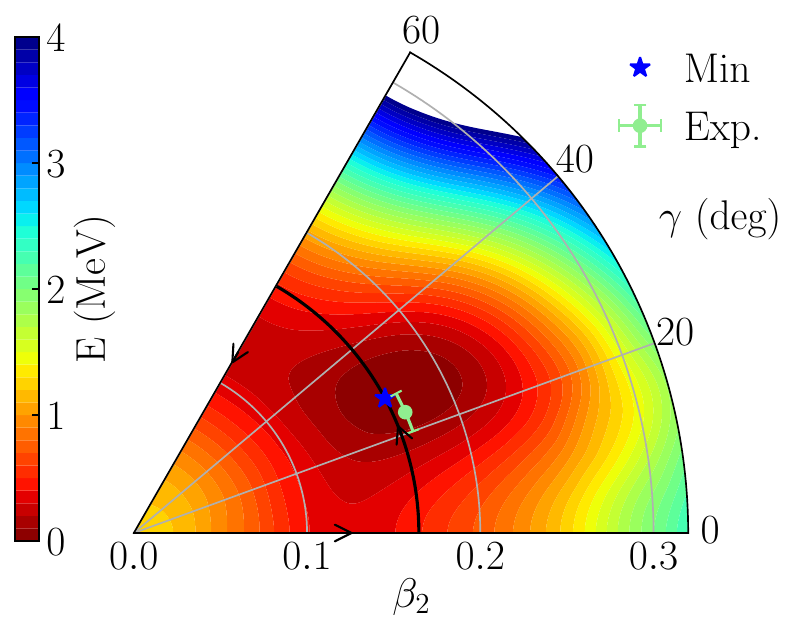}
 \includegraphics[scale=0.3]{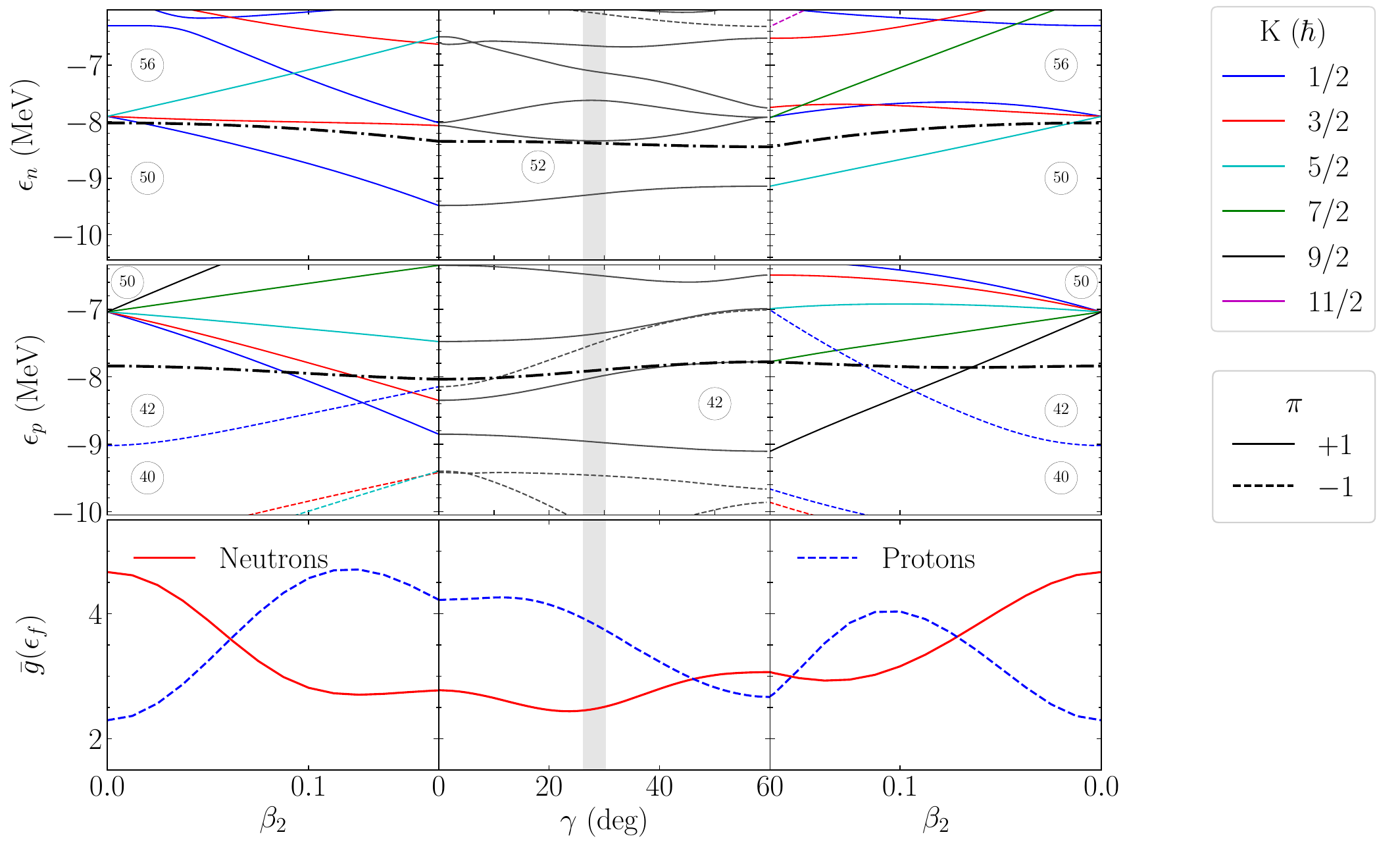}
 \caption{Left: total energy of $^{96}$Mo (normalized to the minimum) in the ($\beta_{2},\gamma$) plane obtained BSkG3; the global minimum as well as the deformation deduced from experiment is indicated. For the latter, we used $\beta_2$ as deduced from measured $B(E2)$ transitions~\cite{Raman01} and $\gamma$ deduced from Coulomb excitation experiments~\cite{svensson1995,Raman01,Zielinska05}. Right: the SPL energies around the Fermi level (dash-dotted black line) along the prolate -- triaxial -- oblate deformation loop marked by the arrows in the ($\beta_{2},\gamma$) plane for neutrons (top panel) and protons (middle panel). The bottom panel
 shows the average SPL density around the Fermi energy (Eq.~\ref{eq_spl}) for neutrons and protons. The gray vertical line shows the location of the global minimum.
 }
 \label{fig_96Mo_pes}
 \end{center}
 \end{figure*}

Throughout the paper, we will use two nuclei as representative examples: the left panels of Fig.~\ref{fig_96Mo_pes} and~\ref{fig_108Pd_pes} show the potential energy surfaces (PESs) in the $\beta_2-\gamma$ plane\footnote{See Appendix~\ref{sec:orientation} for the precise definition of $\beta_{2}$ and $\gamma$. } for, respectively, $^{96}$Mo and $^{108}$Pd. The predicted mean-field minima - indicated by blue stars - are triaxial in both cases; $\gamma \approx 30^{\circ}$ and $\gamma \approx 20^{\circ}$ for the Mo and Pd isotope respectively. For both isotopes, values for $\beta_2$ and $\gamma$ related to rotational invariants deduced from Coulomb excitation experiments exist: these are indicated on the plots as pale green dots. Although this comparison has to be treated with care~\cite{poves2020}, the match between the multipole moments of the BSkG3 ground state and these experimental values is encouraging. There are two important differences between both isotopes for our discussion. First,  the PES of $^{96}$Mo is soft, with the triaxial minimum only about 300 keV below the prolate and oblate saddle points, while that of $^{108}$Pd is more rigid. Second, the total quadrupole deformation of $^{108}$Pd with $\beta_{2} \approx 0.28$ is significantly larger than that of $^{96}$Mo corresponding to $\beta_{2} \approx 0.16$.

%------------------------------------------------

 \begin{figure*}
 \begin{center}
 \includegraphics[scale=0.45]{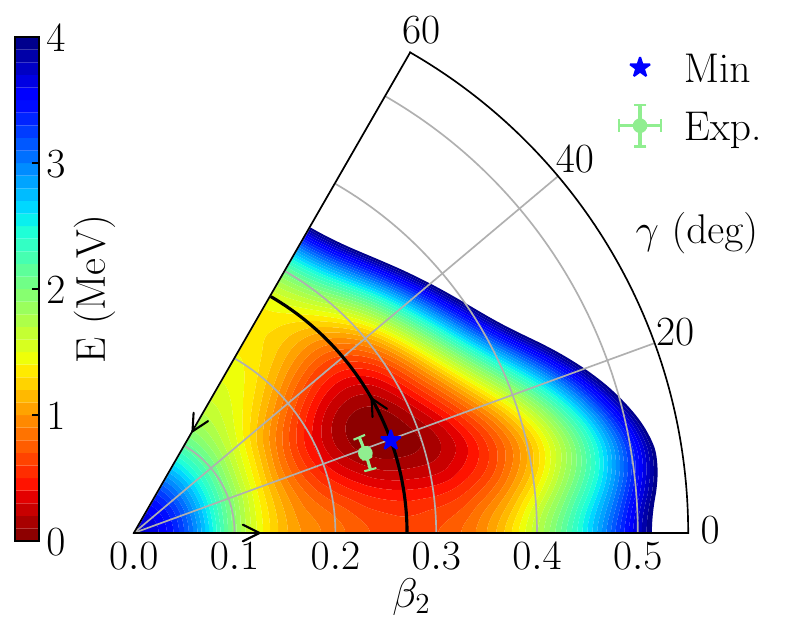}
 \includegraphics[scale=0.3]{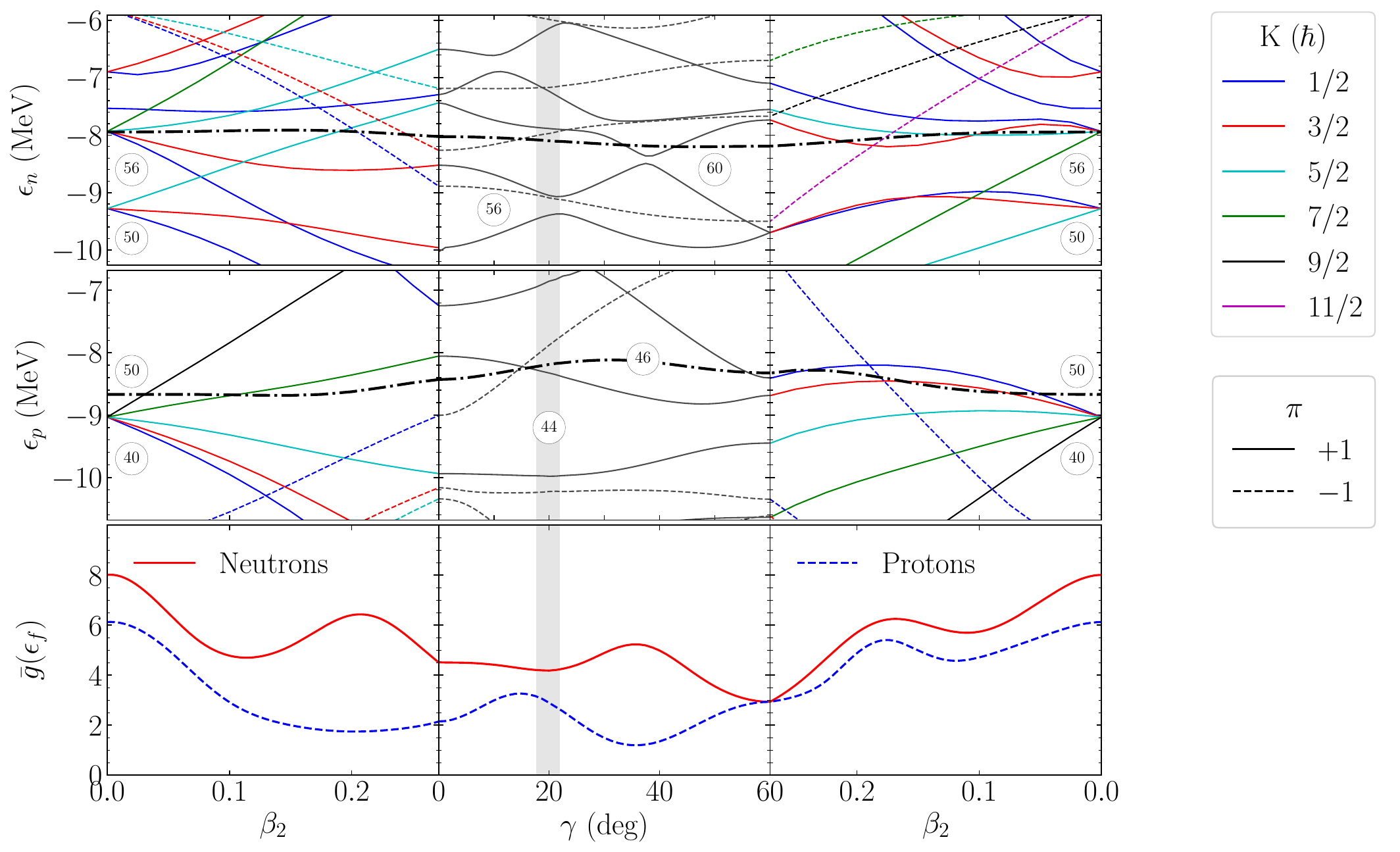}
 \caption{Same as Fig.~\ref{fig_96Mo_pes} but for $^{108}$Pd. For the indicated experimental information we used $\beta_2$ as deduced from measured $B(E2)$ transitions~\cite{Raman01} and $\gamma$ deduced from Coulomb excitation experiments~\cite{svensson1995}.
 }
 \label{fig_108Pd_pes}
 \end{center}
 \end{figure*}

%%%%%%%%%%%%%%%%%%%%%%%%%%%%%%%%%%%%%%%%%%%%%%%%%%%%%%%%%%%%%%%%%%%%%%%%%%%%%%ù
\subsection{Single-particle spectra and intrinsic level densities}
\label{sect_intrinsic}

 The most important ingredient of the combinatorial approach to NLD is the
 single-particle spectrum; from the eigenvalues $\epsilon_{q, i}$ ($q = p/n$) of
 the neutron/proton single-particle hamiltonian $h_{q}$ one can start constructing
 PH excitations and estimate their excitation energy with respect
 to the mean-field mininum~\cite{Hilaire01}. The SPL directly reflect the properties
 of the mean-field minimum; to illustrate this point in general and the consequences of triaxial deformation in particular, the right panels of Figs.~\ref{fig_96Mo_pes}-\ref{fig_108Pd_pes} show the details of the SPL scheme for the two example nuclei along the paths indicated on the corresponding PESs (left panels).

Starting from the spherical point and moving right on the PES, the spherical degeneracy is limited with increasing $\beta_{2}$; single-particle states lose their spherical quantum numbers, but they can be labelled by $K$ - indicated by colors -  the projection of their angular momentum on the remaining rotational symmetry axis. As soon as $\gamma$ differs from $0^{\circ}$ and $60^{\circ}$, $K$ loses its meaning too and parity - indicated by full ($\pi=+$) and dashed ($\pi=-$) lines - is the only quantum
number that we can use to label the states in Fig.~\ref{fig_96Mo_pes} and \ref{fig_108Pd_pes}\footnote{There is another quantum number in our calculations, z-signature $\eta = \pm i$, that can be used to distinguish between time-reversal states. Since time-reversal is conserved in these calculations, all lines on the right panels of Figs.~\ref{fig_96Mo_pes} and \ref{fig_108Pd_pes} are two-fold degenerate and $\eta$ is not useful as a label.}. The evolution of the SPL as a function of $\gamma$ is non-trivial: levels starting on the prolate side with specific $K$ mix to form
different shell gaps at finite $\gamma$, before connecting on the oblate side to
states with different $K$ but linked to the same shell at the spherical point. The shell effects
induced by changes in $\gamma$ are the ultimate origin of the appearance of triaxial deformation; the most apparent example in these figures is the opening of the large $Z=44/46$ proton gaps for $^{108}$Pd that have been suggested as the primary drivers of triaxial deformation in this region of the nuclear chart~\cite{hukkanen2023}. Although we do not show any explicit examples, similar considerations apply to mean-field configurations that break reflection symmetry: the SPL lose the parity quantum number $\pi$ and evolve in a non-trivial way as a function of octupole deformation, typically resulting in shell gaps opening up at so-called octupole magic numbers~\cite{butler1996,chen2021}.

Aside from the detailed evolution of the individual SPL, we also illustrate the evolution of the single-particle density of levels around the Fermi energy: in the bottom panels of the single-particle diagrams in Figs.~\ref{fig_96Mo_pes} and \ref{fig_108Pd_pes} we plot the average SPL density $\bar{g}$ for each isospin $q = p/n$, defined as
\begin{equation}
\bar{g}(\epsilon_{fq}) = \frac{1}{\sqrt{2\pi}\sigma_f} \int_{-\infty}^\infty f_{q,\rm spl} (\epsilon)~\exp\left[-{\frac{(\epsilon-\epsilon_f)^2}{2\sigma_f^2}}\right]  d\epsilon \quad,
\label{eq_spl}
\end{equation}
where $f_{q, \rm spl}(\epsilon) \equiv \sum_{i} \delta(\epsilon - \epsilon_{q,i})$ is the discrete level density in the calculated spectrum and $\sigma_f=0.5$~MeV is adopted. It is clear that $\bar{g}$ varies significantly as a function of both $\beta_2$ and $\gamma$, particularly when remembering that the average level density around the Fermi energy determines the intrinsic NLD exponentially in the simplified picture of an ideal Fermi gas~\cite{Bethe36}. We note in particular that for our open-shell example nuclei the average SPL density for deformed states is much lower than that of the spherical configurations. Because of Strutinsky's theorem, the mean-field energy minimum can be found at the deformation where protons and neutrons can compromise on achieving the lowest possible average single-particle density around their respective Fermi energy~\cite{strutinsky1968}. As can be guessed from Fig.~\ref{fig_108Pd_pes}, our calculations slightly violate Strutinsky's theorem because BSkG3 include energy contributions simulating beyond-mean-field effects - such as a correction for spurious rotational motion - that are not treated variationally. Nevertheless, one should expect that breaking axial symmetry will usually lead to a \emph{decrease} in the average single-particle density around the Fermi energy and hence to a \emph{decrease} in the intrinsic level density as compared to axial configurations of the same nucleus. This is entirely analogous to to the breaking of spherical symmetry: for open-shell nuclei - i.e. the vast majority of all nuclei - the intrinsic level density for spherical configurations is much larger than that of prolate or oblate states.

Past applications of the combinatorial approach started by considering incoherent PH excitations based on single-particle spectra like those in the axially symmetric parts of the level schemes in Figs.~\ref{fig_96Mo_pes} and \ref{fig_108Pd_pes}; this way one can construct the \emph{intrinsic} state density $\rho_i(U,K,\pi)$ that depends on (i) the excitation energy $U$, (ii) the angular momentum quantum number $K$ and (iii) the parity $\pi$. We cannot proceed in the same way here as many nuclei have predicted ground states that feature either triaxial or octupole deformation
\footnote{Remarkably, BSkG3 does not predict any ground states that combine triaxial and octupole deformation~\cite{Grams23}.}; hence for many nuclei we have no access to either $K$ or $\pi$.

We could forego both of these quantum numbers and construct the intrinsic level density $\rho_i(U)$ purely as a function of excitation energy, but this would forbid us from constructing the angular-momentum and parity-dependent NLDs that are one of the main advantages of the combinatorial NLD approach. To circumvent this issue, we start by defining
a \emph{rounded} $\bar{K}$ quantum number for each single-particle state $\psi$ as
\begin{equation}
\bar{K} = \frac{1}{2} \lfloor 2 \langle \psi | \hat{J}_{\mu} | \psi \rangle \rceil\, ,
\label{eq:barK}
\end{equation}
where $\lfloor \rceil$ indicates a rounding operation to the nearest integer and $\hat{J}_{\mu}$ is the single-particle angular momentum operator around a reference axis $\mu$. The latter needs to be chosen appropriately, a non-trivial task when dealing with a triaxial shape that has three distinct principal axes; in practice we choose $\mu = z$ and orient the nucleus in our three-dimensional simulation volume such that the Belyaev rotational MOI along the z-axis is the smallest, see App.~\ref{sec:orientation} for more details. This definition of the reference axis ensures that we recover the correct limit when the mean-field configuration is axially symmetric: in that case $\bar{K}$ reduces to the quantum number $K$ for both prolate and oblate configurations as the MOI associated with a rotational symmetry axis vanishes.

With Eq.~\eqref{eq:barK} in hand, we can construct an intrinsic state density that depends on excitation energy and $\bar{K}$, i.e. $\rho_i(U,\bar{K})$. To recover a parity quantum number for octupole deformed states, we assume strong symmetry breaking, such that each state without left-right symmetry corresponds to \emph{two} states with opposite parity at identical excitation energy, i.e. that $ \rho_i(U,\bar{K},\pi=+) = \rho_i(U,\bar{K},\pi=-) = \rho_i(U,\bar{K})$. While our prescription for triaxial deformation is a new proposal, this strategy to deal with octupole deformation dates back to Ref.~\cite{bjornholm1974} and has already been used in combinatorial calculation for saddle point NLDs in the context of fission \cite{Goriely08b}; for this reason we focus our examples on the treatment of nuclei with triaxial deformation.

Proceeding in this way we recover an intrinsic state density that depends on excitation energy, angular momentum and parity, $\rho_i(U, \bar{K}, \pi)$ at the cost of some rather ad-hoc approximations. We can expect for instance that
$\bar{K}$ will somewhat reliably single out the dominant $K$ component of a single-particle state if $\gamma$ is close
to $0^{\circ}$ or $60^{\circ}$, but this is certainly not true for configurations with more clearly developed triaxial deformation. In a similar vein, the equipartition of parity we enforce for octupole deformation is
only valid for heavy nuclei with large octupole deformation~\cite{robledo2011}; in general one expects a quite significant energy difference between the two states of opposing parity once projected.
Unfortunately, truly disentangling complex symmetry-broken states into components with associated quantum numbers requires advanced many-body techniques that are not feasible on the scale required for global NLD calculations today. Despite these deficiencies, our approach in general and Eq.~\eqref{eq:barK} in particular  does not introduce any new parameter and conserves the (limited) physical information that can be gleaned from the single-particle expectation values of the angular momentum operators.

%%%%%%%%%%%%%%%%%%%%%%%%%%%%%%%%%%%%%%%%%%%%%%%%%%%%%%%%%%%%%%%%%%%%%%%ùù

\subsection{Collective motion}
\label{sect_collective}

The density of PH excitations cannot on its own account for the low-energy spectroscopy of nuclei; we further include corrections for rotational and vibrational motion. To account for the latter, we fold the incoherent PH excitations with the phonon state densities as described by a boson partition function~\cite{Hilaire01}. Unfortunately, we are not yet capable of building such partition functions consistently from BSkG3 calculations; instead we use experimental information when available and the shell-dependent analytical expression of Ref.~\cite{Goriely08b} to estimate the excitation energies of vibrational modes. As in Ref.~\cite{Goriely08b},
a maximum number of three phonons is considered and all the quadrupole, octupole and hexadecapole vibrational modes are included. To account for the damping of vibrational effects at increasing excitation energy, we restrict the folding to the ph configurations having a total exciton number (i.e. the sum of the number of proton and neutron particles and proton and neutron holes) $N_{\rm ph} \leq 4$.
This restriction stems from the fact that a vibrational state results from a coherent excitation of particles and holes, and that this coherence vanishes with increasing number of ph involved in the description. Therefore, if one deals with a ph configuration having a large exciton number, one should not simultaneously account for vibrational states which are clearly already included as incoherent excitations. Once the vibrational and incoherent PH state densities are computed, they are folded to deduce the total state densities $\rho(U,\bar{K},\pi)$.

We include rotational collectivity by explicitly constructing the rotational bands on top of each of the intrinsic states. More specifically, we use the Hamiltonian of a rigid (and possibly asymmetric) rotor to build rotational spectra:
\begin{equation}
\hat{H}_{\rm rot} = \sum_{\mu= x,y,z} \frac{\hat{J}^2_{\mu}}{2 \mathcal{I}_{\mu}} \, ,
\label{eq_hrot}
\end{equation}
where $\mathcal{I}_{x/y/z}$ are the MOIs of the ground state; we assume these MOIs are representative of all band heads in the same nucleus. We use the Belyaev
MOI but rescale it with a factor of 1.32; this simple procedure has turned out to be an excellent approximation to the much more involved calculation of a Thouless-Valatin MOI, which
- contrary to a Belyaev MOI - does account for the response of nuclear mean fields to rotation~\cite{ryssens23}. As shown in Ref.~\cite{Grams23}, BSkG3 predictions for the
MOIs of medium to heavy nuclei across the nuclear chart are more than reasonable with the exception of an underestimation in the actinide region; BSkG4 rectifies this issue and
the influence on NLD calculations will be investigated elsewhere~\cite{grams2025}.

Numerically diagonalising Eq.~\eqref{eq_hrot} leads to a spectrum of rotational states labelled by an integer rotational quantum number $J_{\rm rot}$ with associated rotational energies $E^{J_{\rm rot}}_{i}$,
where $i$ ranges from $1$ to $J_{\rm rot} + 1$ ($J_{\rm rot}$) if $J_{\rm rot}$ is even (odd)~\cite{ray32}. In the axially symmetric case the coupling between intrinsic angular momentum $K$ and the collective angular momentum $J_{\rm rot}$ is straightforward because the former aligns along the symmetry axis while the latter is perpendicular to the symmetry axis. Unfortunately, the angular momenta are no longer confined in this way when the nucleus is triaxially deformed and we are not aware of any way to rigorously perform the angular momentum coupling that is feasible for global NLD calculations. Instead, we adopt the simplest possible recipe that does not discard any information inherent in the values of $J_{\rm rot}$ and $\bar{K}$: we assume that both angular momenta add coherently, i.e. we assign a total angular momentum $  J \equiv J_{\rm rot} + | \bar{K} |$ to a configuration coupling an intrinsic state characterized by $\bar{K}$ that rotates collectively with angular momentum $J_{\rm rot}$.

With this simple coupling scheme, we are now in a position to build the level density of a triaxially deformed nucleus:
\begin{equation}
\rho(U, J, \pi) = \frac{1}{2-\delta_{\bar{K},0}}\sum_{\bar{K} = -J}^{J} \sum_{i=1}^{n(J,|\bar{K}|)} \rho_i \left(U - E_i^{J - |\bar{K}|},\bar{K}, \pi\right) \,,
\label{eq:deformed}
\end{equation}
where the prefactor accounts for states with opposite $\bar{K}$ leading to the same rotational levels. The second sum in this equation ranges over all members of the multiplet of the rotational Hamiltonian characterized by $J_{\rm rot} = J - |\bar{K}|$ that are physically allowed; we indicate their number by $n(J,|K|)$. It is important to note that the latter is \emph{not} equal to $2J_{\rm rot}+1$; as symmetry considerations imply that not all collective wavefunctions can be coupled with intrinsic states with $\bar{K} = 0$~\cite{younes2004,rowe2010};
more specifically, we have 
\begin{equation}
n^{J, |\bar{K}|} =
\left\{
\begin{matrix}
& J_{\rm rot}+1   & \text{ if } J_{\rm rot} \text{ is even, and } |\bar{K}| \not= 0,\\
& J_{\rm rot}     & \text{ if } J_{\rm rot} \text{ is odd, and } |\bar{K}| \not= 0,\\
& J_{\rm rot}/2+1 & \text{ if } J_{\rm rot} \text{ is even, and } |\bar{K}| = 0,\\
& (J_{\rm rot}-1)/2   & \text{ if } J_{\rm rot} \text{ is odd, and } |\bar{K}| = 0.\\
\end{matrix}
\right.
\end{equation}
For a triaxially deformed nucleus, all of the $n^{J, |\bar{K}|}$ states will have rotational energies of at most a few MeV's. An axially symmetric nucleus could hypothetically be
treated in the same way; in that case one of the $\mathcal{I}_{\mu}$ would vanish and all except one of the rotational states for a given $J$ would rise to infinite excitation energy~\cite{davydov1958}.
For this reason, we can expect the NLD of a triaxial nucleus to be larger than that of one that retains axial symmetry, \emph{provided the intrinsic level densities are equal}.

\begin{figure}[htbp]
\begin{center}
\includegraphics[scale=0.5]{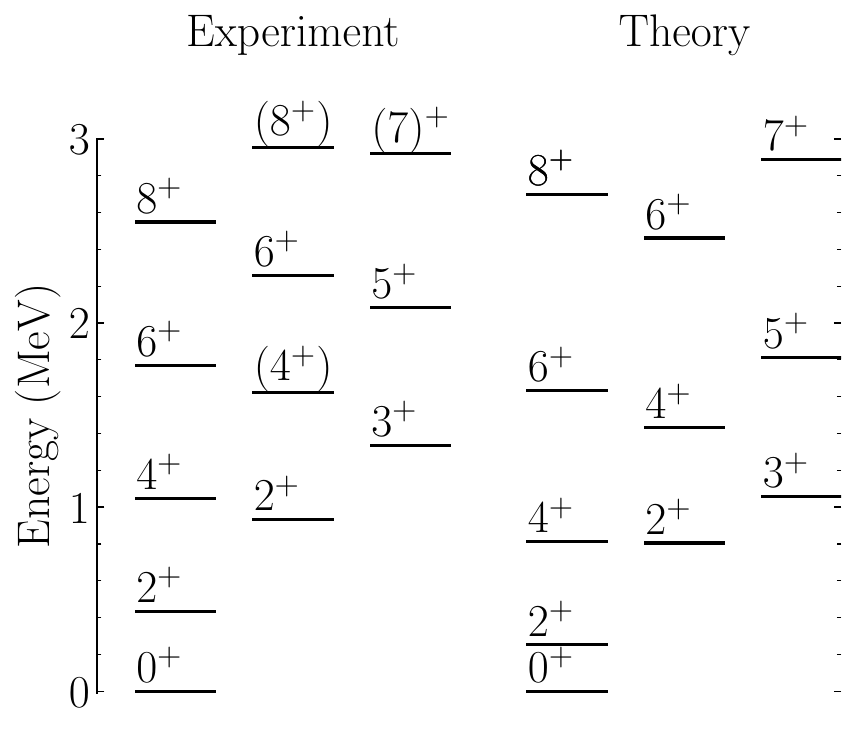}
\caption{Comparison between the experimental (left) and theoretical (right) rotational energy spectrum of the triaxial nucleus $^{108}$Pd.}
\label{fig_108Pd_spectrum}
\end{center}
\end{figure}

As an example, we compare the rotational spectrum constructed in this way on top of the mean-field ground state in Fig.~\ref{fig_108Pd_spectrum} with the relevant part of the
known experimental spectrum. The reproduction of the ground-state band and the triaxial side-bands is not perfect but certainly satisfactory in light of our simplistic assumption
of rigid rotation. This figure also illustrates the effect of triaxial deformation on the collective rotational enhancement: the $2^+, 3^+, 4^+, \ldots$ sequence that is characteristic of triaxial deformation would be absent in an axially symmetric nucleus. We do not show an equivalent figure for $^{96}$Mo, since its spectrum is known not to be a rotational one, as reflected by the softness of the BSkG3 PES and the smallness of the calculated MOIs.

When the calculated nuclear ground state retains axial symmetry, we recover $K$ and can characterize the nuclear rotation by a single MOI $\mathcal{I}_{\perp}$; we then revert
to the well-known formula for the rotational enhancement~\cite{Hilaire01}
\begin{eqnarray}
\rho_{\rm def}(U,J,\pi) & = & \displaystyle{
\frac{1}{2} \,\,
\Bigl[
\sum_{K=-J,K \neq 0}^{J}} \,\,\,\,\, \rho_{i}(U-E_{\rm rot}^{J,K},K,\pi)
\Bigr] \nonumber \\
& + & \delta_{(J \rm \,=even)} \, \delta_{(\pi=+)} \,
\rho_{i}(U \! - \! E_{\rm rot}^{J,0},0,\pi) \nonumber \\
& + & \delta_{(J \rm \,= odd)} \, \delta_{(\pi=-)} \,
\rho_{i}(U \! - \! E_{\rm rot}^{J,0},0,\pi) \mbox{,}
 \label{rotmic} \\
{\rm where} \quad  E^{\rm J,K}_{\rm rot} &=& \frac{J(J+1) - K^2}{2 \mathcal{I}_{\perp}} \, .
\end{eqnarray}

If instead the nucleus is predicted to be spherical, the intrinsic and laboratory frames coincide, and the level density is trivially obtained through the relation
\begin{equation}
\rho_{\rm sph}(U,J,\pi)=\rho_{i}(U,M \!\! = \!\! J,\pi)-\rho_{i}(U,M \!\! =
\!\! J \! + \! 1,\pi) \mbox{.}
\label{sphmic}
\end{equation}
%%%%%%%%%%%%%%%%%%%%%%%%%%%%%%%%%%%%%%%%%%%%%%%%%%%%%%%%%%%%%%%%%%%%%%%%%%%%%%%%%%
\subsection{ Weak symmetry breaking }
\label{sect_damping}

By combining Eqs.~\eqref{eq:deformed}, \eqref{rotmic} and \eqref{sphmic}, we are suitably equipped for the description of spherical nuclei and well-deformed nuclei, whether
they retain a rotational symmetry axis or not. The intermediate case - weakly deformed nuclei with correspondingly small correlation energies - is problematic as simple models
such as those invoked above do not apply; in particular, the final NLD results will depend in a discontinuous way on whether a nucleus is classified as `spherical', `axially symmetric'
or `triaxial'.

For a deformed nucleus, we judge whether it is axially symmetric or not based on its calculated MOIs: if $\min (\mathcal{I}_x, \mathcal{I}_y, \mathcal{I}_z) < 0.1 $ MeV$\hbar^{-2}$, then we
label it as axially symmetric. In exploratory calculations, we found that the difference between using Eq.~\eqref{eq:deformed} and Eq.~\eqref{rotmic} is typically limited to a factor of about 50\% (an example is given for the outer barrier in Fig.~\ref{fig_nld_239U_Bf_vs_GS}); given the expected accuracy of a global NLD model we opted to accept this modest discontinuity instead of introducing a phenomenological smoothing between both regimes.

The case is different for the transition between spherical and deformed where the discontinuity is much larger; following earlier studies~\cite{Goriely08b,Hilaire06}, we introduce a damping function ${\cal F}$ that is now a function of $r_{42}=E_1(4^+)/E_1(2^+)$, a quantity that should be equal to 3.33 for a well deformed rotational nucleus and a value of 2.0 for a vibrator. The damping function reads
\begin{equation}
{\cal F}=1-\Bigl[1+{\rm e}^{\displaystyle(r_{42}-r_{42}^*)/d_r}\Bigr]^{-1},
\label{eq:fdam}
\end{equation}
so that the final NLD is obtained through
\begin{equation}
\rho(U,J,\pi)=\Bigl[1-{\cal F} \Bigr] \rho_{\rm sph}(U,J,\pi) + {\cal F} \rho_{\rm def}(U,J,\pi).
\label{eq:fdam2}
\end{equation}
The $r_{42}$ parameter are taken from the large-scale calculation of Ref.~\cite{Hilaire07} on the basis of the Generalized Coordinate Method with the D1M Gogny interaction. The parameters $r_{42}^* = 3.0$ and $d_r = 0.4$ in Eq.~\eqref{eq:fdam} have been adjusted in order to reproduce at best the measured s-wave mean spacings on the basis of the BSkG3 NLD (see Sect.~\ref{sect_res}). Future developments will allow us to consistently determine $r_{42}$ on the basis of BSkG3.
Note that Eqs.~\ref{eq:fdam}-\ref{eq:fdam2} do not describe the disappearance of rotational effects with increasing excitation energies, as predicted by different theoretical works \cite{Hansen83,Martin03,Hilaire12}. It is possible to include such effects in a combinatorial approach through temperature-dependent HFB calculations \cite{Hilaire12};  we do not do so here because of the numerical burden that would pose for our three-dimensional numerical representation.

%%%%%%%%%%%%%%%%%%%%%%%%%%%%%%%%%%%%%%%%%%%%%%%%%%%%%%%%%%%%%%%%%%%%%%%%%%%%%%%%%%%%%%%%%%%%%%
\subsection{ The NLD of $^{96}$Mo and $^{108}$Pd }
\begin{figure}[htbp]
 \begin{center}
 \includegraphics[scale=0.45]{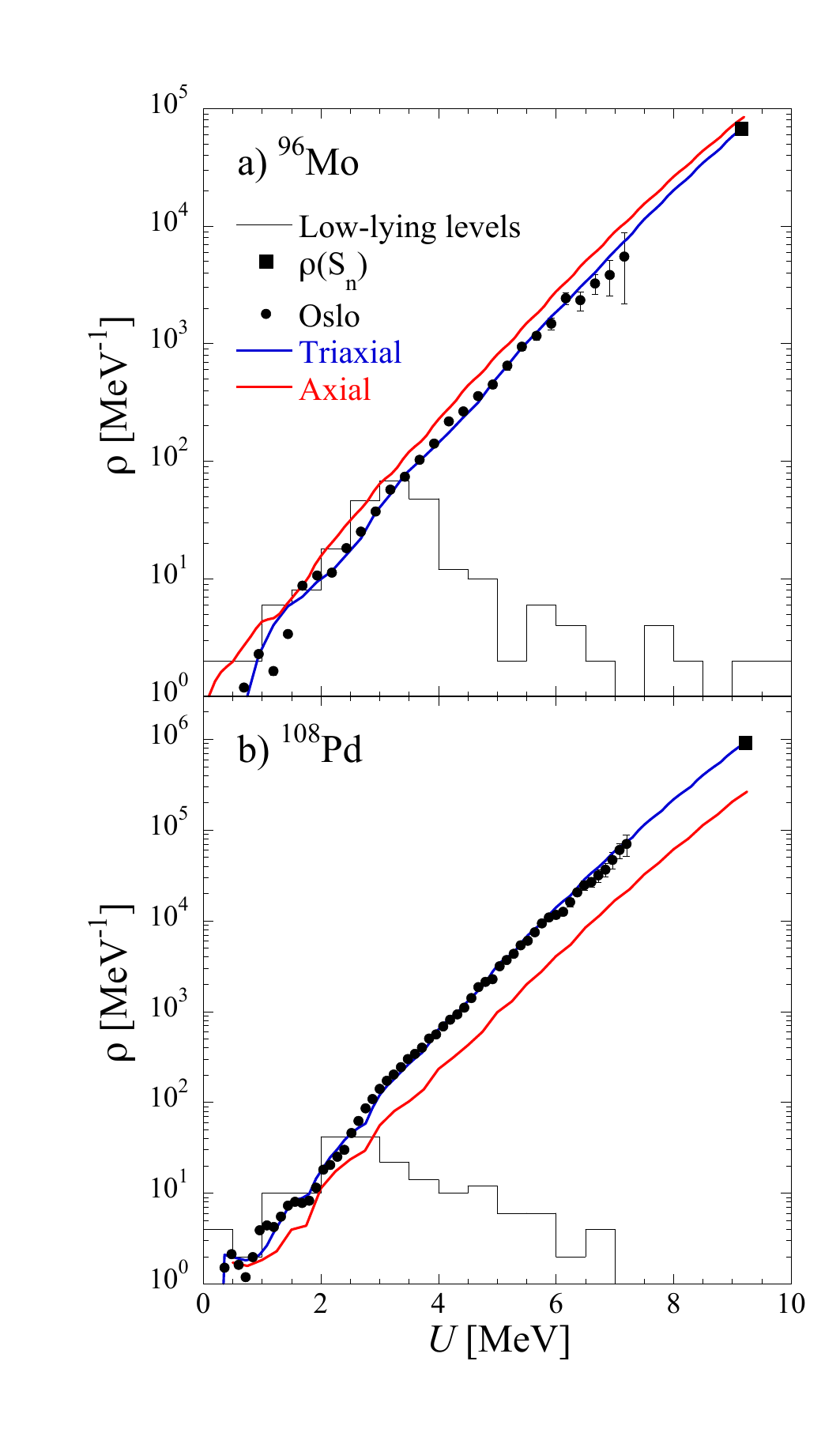}
 \vskip -0.5cm
 \caption{ Illustration of the impact of triaxiality on the total NLD of a) moderately deformed $^{96}$Mo ($\beta_{20}=0.14$,  $\gamma=28^{\circ}$) and b) well deformed $^{108}$Pd ($\beta_{20}=0.26$,  $\gamma=20^{\circ}$); see text for details.
   %The blue curves are obtained taking into account the triaxial deformation of the ground state, while the red curve considers the axial deformation with the lowest energy. For each nucleus, both curves have been renormalized by the same $\alpha$ and $\delta$ parameters (see Eq.~\ref{eq_renorm}), so that the triaxial calculation reproduces the total level density at $U=S_n$ from experimental s-wave spacings. The black line shows the density extracted from known low-lying levels \cite{Capote09}, the black squares the total level density derived from the known s-wave spacings \cite{Capote09} and the black circles the Oslo data \cite{Oslo} renormalized as detailed in Ref.~\cite{Goriely22a}.
 }
 \label{fig_Mo_Pd}
 \end{center}
 \end{figure}

Fig.~\ref{fig_Mo_Pd} illustrates the final calculated NLD of $^{96}$Mo and $^{108}$Pd, renormalized to the available experimental data as explained below around Eq.~\eqref{eq_renorm}. Our final results that account for triaxial deformation are displayed as blue lines, while the NLD obtained from applying our modelling - including identical renormalization - to the axially symmetric saddle point (oblate for $^{96}$Mo, prolate for $^{108}$Pd) are given in red.
%SG: this is correct
The plots also include the experimental level density extracted from (i) known low-lying levels~\cite{Capote09}, (ii)  the measured s-wave resonance spacings~\cite{Capote09} at the neutron separation energy $U = S_n$, and (iii) Oslo data~\cite{Oslo} renormalized on the total level density at $S_n$ obtained with the triaxial calculation, as detailed in Ref.~\cite{Goriely22a}.

Fig.~\ref{fig_Mo_Pd} showcases the overall impact of triaxial deformation: accounting for this exotic deformation can - depending on the nucleus - either decrease ($^{96}$Mo) or increase ($^{108}$Pd) the calculated NLDs compared to axially symmetric calculations. In our microscopic framework, this results from a competition between two effects: as compared to an axially symmetric calculation, (i) triaxial deformation tends to lower the SPL density around the Fermi energy, leading to smaller intrinsic state densities while (ii) the additional possibilities for collective rotation leads to a larger collective enhancement. The precise balance between both effects will depend on the nucleus: for moderately deformed $^{96}$Mo the reduction in SPL density around the Fermi energy in Fig.~\ref{fig_96Mo_pes} overpowers the moderate difference in between triaxial and axial rotational enhancement; the situation is exactly opposite for $^{108}$Pd. 
%For both $^{96}$Mo and $^{108}$Pd, the triaxial NLD calculations reproduce both the low-lying states and the energy dependence of the Oslo data rather well.

 %------------------------------------------------
\begin{figure}
 \begin{center}
 \includegraphics[width=.4\textwidth]{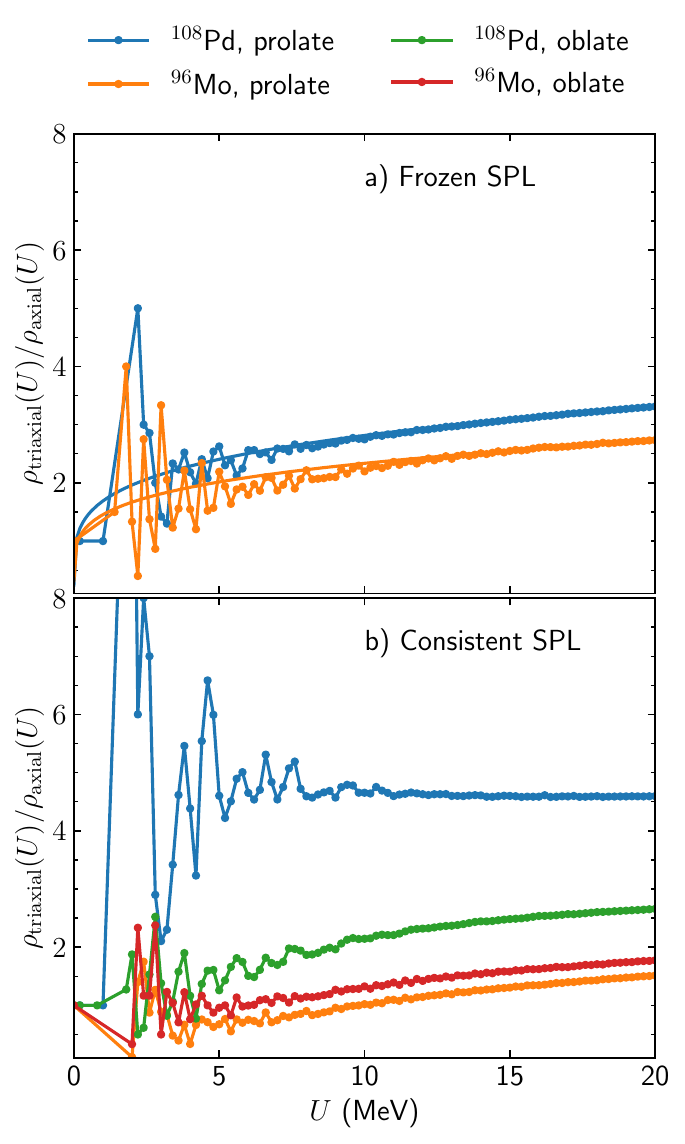}
 \caption{ Energy dependence of the ratio between the NLD based on a triaxial configuration - $\gamma = 20^{\circ}$ and $30^{\circ}$ for for $^{108}$Pd and $^{96}$Mo, respectively - and either the prolate or oblate reference configuration at equal $\beta_2$ without the folding of vibrational bosons.
           a) results for frozen SPL (see text). Only the curves for the prolate reference configuration are shown; taking the oblate
            configuration as reference results in nearly indistinguishable curves. The solid lines are best fits of the $U^{1/4}$ trend to guide the eye.
            b) results for the SPL consistently obtained for the different configurations.}
 \label{fig_Edep}
 \end{center}
 \end{figure}

%--------------------------------------------------
%
To illustrate these competing effects in more detail, we compare two sets of NLD calculations obtained with different single-particle (s.p.) level scheme: a set with `consistent SPL' and one with `frozen SPL'. The former differs only from our final calculation through our omission of vibrational folding for simplicity; it relies on the SPL of the triaxial minimum and the corresponding collective rotational enhancement. The `frozen SPL' calculation also omits the vibrational folding, but applies the triaxial collective enhancement to the SPL of an axially symmetric saddle point on the PES. Fig.~\ref{fig_Edep} shows the enhancement of the resulting NLDs over those obtained assuming axial symmetry: the `frozen SPL' calculation in the top panel and the `consistent SPL' in the bottom panel. In the `frozen SPL' case, we find a $U^{1/4}$ dependence (illustrated by the full lines) at high excitation energies; this matches the dependence one would have expected from analytical formulae~\cite{Capote09,Bjornholm73}. More specifically, one finds essentially that $\rho_{\rm axial}/\rho_i \propto \mathcal{I}_\perp~U^{1/2}$  while $\rho_{\rm triaxial} / \rho_i \propto \sqrt{\mathcal{I}_x \mathcal{I}_y \mathcal{I}_z}~U^{3/4}$; \emph{provided the intrinsic level density $\rho_i$ does not change,} one should expect $\rho_{\rm triaxial}/\rho_{\rm axial} \propto \sqrt{\mathcal{I}_x \mathcal{I}_y \mathcal{I}_z} / \mathcal{I}_\perp \times U^{1/4}$ at large excitation energies, when pairing and shell effects have been washed out. Because analytical formulae only include the collective enhancement, they lead to the prediction that triaxial deformation always enlarges the NLD. However, such non-microscopic approaches cannot capture the evolution of the SPL density as a function of deformation; if we base our calculations on `consistent SPL', we do not systematically recover the $U^{1/4}$ dependence as shown by the `$^{108}$Pd, prolate' curve on the bottom panel of Fig.~\ref{fig_Edep}, nor do we always find an enhancement of the NLD as shown by both $^{96}$Mo curves.

\begin{figure}
%   \begin{center}
  \includegraphics[width=.9\columnwidth]{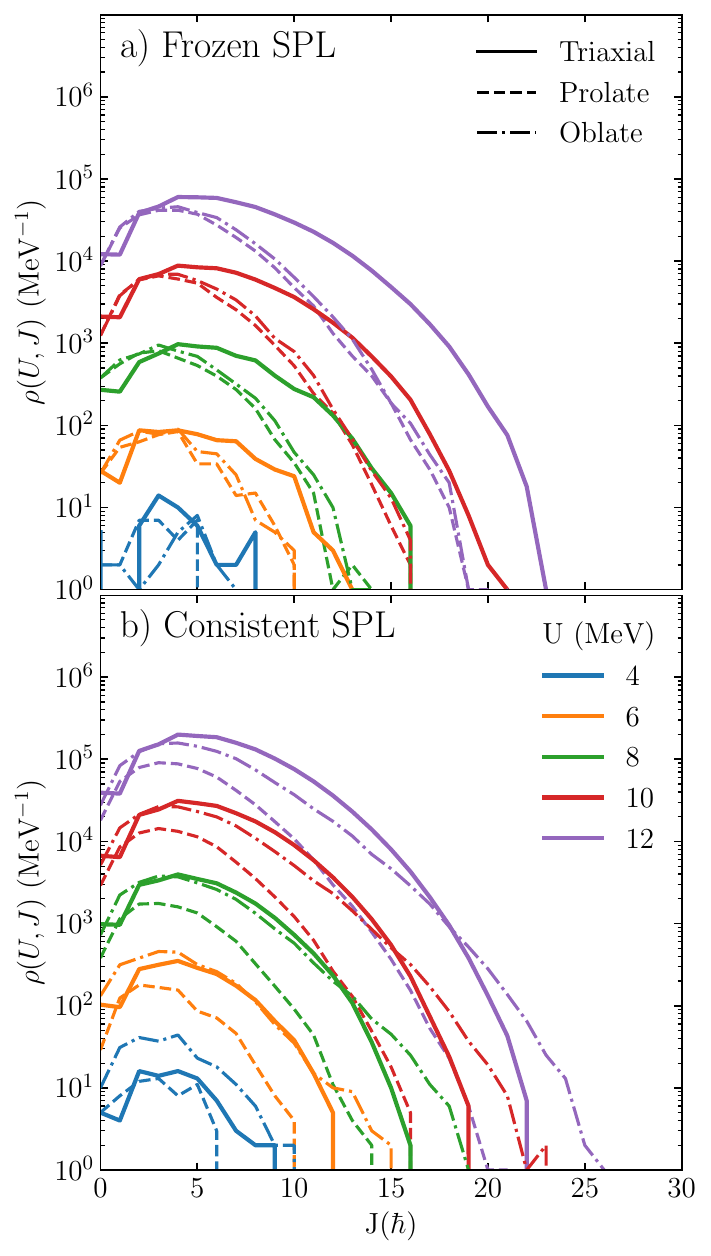}
  \caption{Spin-dependence of the level density of $^{108}$Pd at 5 different excitation energies $U$ between 4 and 12~MeV based on
           three configurations, namely the triaxial minimum (solid lines), the prolate (dashed lines) and oblate (dot-dash lines) axially symmetric saddle points. a) $\rho(U,J)$ obtained from the SPL frozen to those of the prolate saddle point, see text.
           b) $\rho(U,J)$ obtained consistently for SPL at the triaxial minimum and both axially symmetric saddle points (see Fig.~\ref{fig_108Pd_pes}).}
  \label{fig_enh_spin}
  \end{figure}

We also investigate the spin distribution of our calculated NLDs: Fig.~\ref{fig_enh_spin} compares the $^{108}$Pd triaxial spin distribution at 5 different energies with the one obtained when considering axial symmetry for both our `frozen SPL' and `consistent SPL' calculations. Clearly significantly wider distributions are found for triaxial nuclei, i.e. at a given energy, many more high spin states are predicted.  
Additionally, Fig.~\ref{fig_enh_spin} clearly shows that the new spin distributions cannot be described by the simple Gaussian form that is traditionally assumed within the standard Fermi gas modelling \cite{Bethe36,Capote09}.

 %------------------------------------------------
%--------------------------------------------------

 %------------------------------------------------

%--------------------------------------------------
 \section{Comparison with experimental data} \label{sect_res}

 The new NLDs are now compared with experimental data. In spite of considerable experimental efforts made to derive NLD, the
 lack of reliable data -- especially over a wide energy range -- constitutes the main problem for NLD models. Neutron resonance spacings at the neutron separation energy $S_n$ extracted from transmission measurements so far provide the most reliable experimental information on NLD for about 300 nuclei. 
In such an experiment a low-energy neutron gets captured by a target nucleus $(Z,A)$
with a ground-state spin-parity $J_t^{\pi_t}$; the partial level density $\rho$ for the accessible final spin-parity $J_f^{\pi_t}$
of the resulting compound system $(Z,A+1)$ at its neutron separation energy $S_n$ can be related to the s-wave resonance spacing $D_0$ by 
\begin{equation}
\frac{1}{D_0}=  \sum_{J_f} \rho(S_n,J_f=|J_t \pm \tfrac{1}{2}|,\pi_t).
\label{eq:d0}
\end{equation}
%
%  For a nucleus $(Z,A + 1)$ resulting from the capture of a low-energy neutron by a
%  target $(Z,A)$,  for an s-wave neutron-resonance experiment, the neutron resonance spacing $D_0$ can be
%  written in terms of the partial level density $\rho$ for the involved spin(s) and parity as
% \begin{equation}
% \frac{1}{D_0}=  \sum_{J_f} \rho(S_n,J_f=|J_t \pm \tfrac{1}{2}|,\pi_t).
% \label{eq:d0}
% \end{equation}
% where $J_t$ is the spin of the target nucleus ground state and $\pi_t$ its parity.
For a target nucleus with $J_t = 0^+$ capturing an s-wave neutron with spin $s=1/2$ and  orbital angular momentum $\ell = 0$, the populated levels in the compound nucleus will have final spin $J_f = 1/2$ and positive parity. 
If $J_t>0$, the levels populated in the capture process have spins $J_f = J_t \pm 1/2$ with positive parity if  $\pi_t = +$, or negative parity if $\pi_t = -$.
Measured at $U = S_n$, $D_0$ is known to be sensitive to shell, pairing and deformation effects.

Additionally, the cumulative number of low-lying levels also provides key constraints on NLD at the lowest energies. Experimental information on NLDs can also be extracted from other sources, such as primary or multi-step cascade $\gamma$-ray spectra, reaction data, isomeric cross section ratios, average radiative widths, or radiative neutron capture cross sections. However, in these cases, the associated NLDs are also sensitive to additional uncertain ingredients (such as the photon strength function) or affected by systematic errors owing to both experimental and model uncertainties.
 
%------------------------------------------------
 \begin{figure}[htbp]
 \begin{center}
 \includegraphics[scale=0.4]{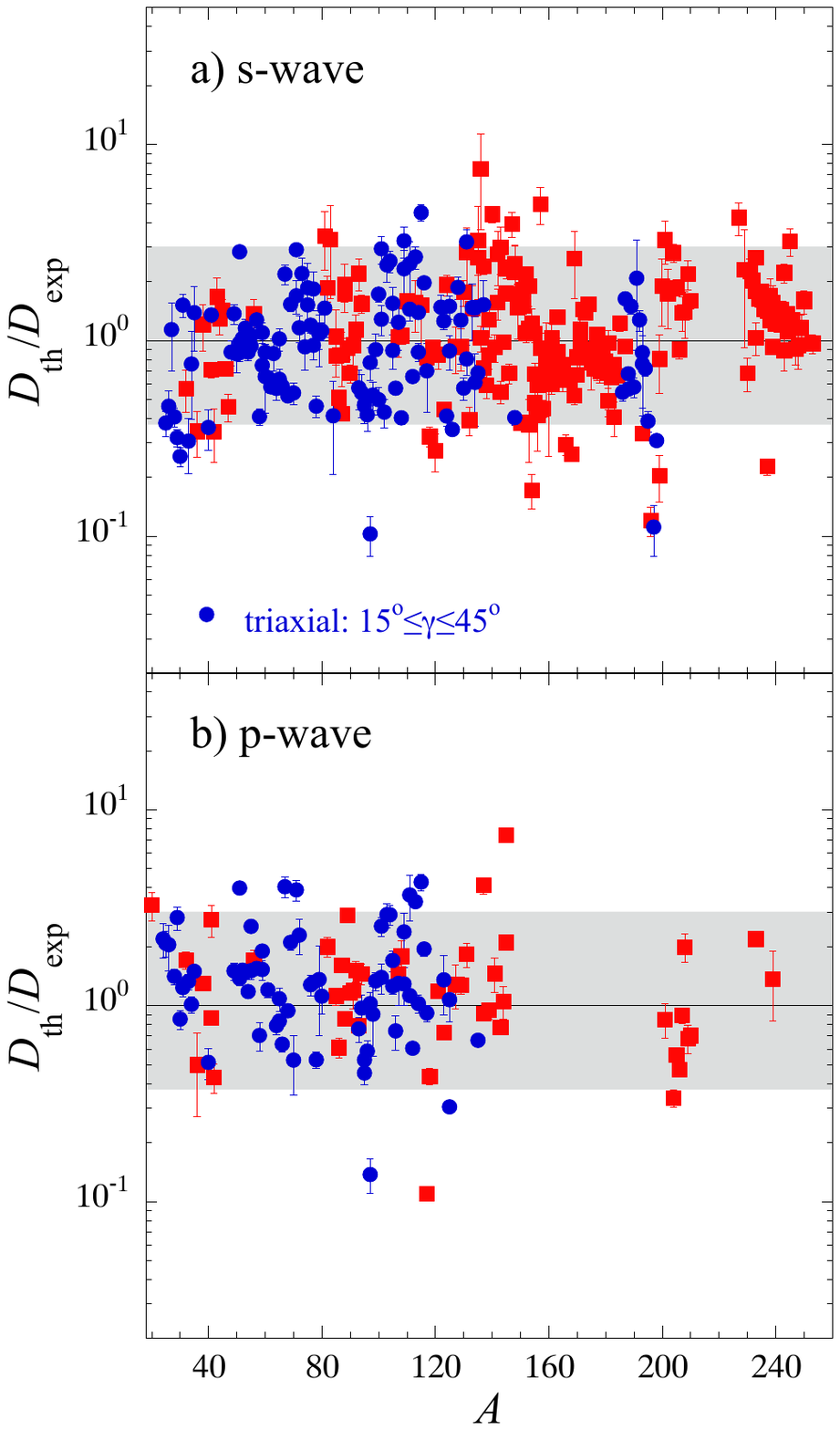}
 \caption{\label{fig_d0} a) Ratio of BSkG3 plus combinatorial ($D_{\rm th}$) to the experimental ($D_{\rm exp}$) s-wave neutron resonance spacings for 299 nuclei compiled in \cite{Capote09}. The blue circles correspond to triaxially deformed nuclei with $15^{\circ} \le \gamma \le 45^{\circ}$, the others are shown with red squares. Error bars correspond to experimental uncertainties on $D_{\rm exp}$. b) same as a) for the p-wave neutron resonance spacings of 116 nuclei  \cite{Capote09}.}
 \end{center}
 \end{figure}
%--------------------------------------------------

%------------------------------------------------
% \begin{figure}
% \begin{center}
% \includegraphics[scale=0.30]{Figures/fig_D0_axial.pdf}
% \caption{\label{fig_d0_axial} Ratio of BSkG3 plus combinatorial ($D_{\rm th}$) to the experimental ($D_{\rm exp}$) s-wave neutron resonance spacings  compiled in \cite{Capote09} for 69 even-even nuclei. The blue circles correspond to the triaxial ground state, if triaxially deformed, and the red squares to ground state assumed to be axially deformed. }
% \end{center}
% \end{figure}
%--------------------------------------------------

 Concerning the neutron resonance spacings, we compare in the top panel of Fig.~\ref{fig_d0} our predictions to the experimental s- and p-wave spacings compiled in the RIPL-3 database~\cite{Capote09}.
 The quality of a global NLD formula can be quantified by the rms deviation factor defined as
 \begin{equation}
f_{\rm rms} = \exp  \left[ \frac{1}{N_e} \sum_i^{N_e} \left( \ln r_i \right) ^2 \right] ^{1/2}
 \label{eq_frms0}
 \end{equation}
and the mean deviation
 \begin{equation}
f_{\rm mean} = \exp  \left[ \frac{1}{N_e} \sum_i^{N_e} \ln r_i \right]
 \label{eq_erms0}
 \end{equation}
 \noindent where $r_i$ is the ratio of theoretical to experimental resonance spacing and $N_e$ is the number of nuclei in the compilation. As detailed in Appendix ~\ref{sect_appendix}, different expressions of the theoretical-to-experimental ratio $r_i$ can be considered to take the experimental uncertainties into account. We adopt here the expression given by Eq.~(\ref{eq_frms2}) which assumes that the experimental data is well described by a normal distribution of width $\delta D^i_{\rm exp}$ (the experimental uncertainty \cite{Capote09}) around the experimental value. Using this indicator, we find that our predictions for 299 s-wave experimental spacings match experiment with an rms deviation $f_{\rm rms}=1.96$ and mean deviation $f_{\rm mean}=1.03$, and the 116 p-wave spacings by $f_{rms}=1.99$ and $f_{\rm mean}=1.21$. This result should be compared to the $f_{rms}=2.34$ deviation of the BSk14 plus combinatorial model \cite{Goriely08b} for s-wave and 2.24 for p-wave spacings. Our new combinatorial model therefore gives a rather improved description of experimental data when compared to our previous calculation. Note that this improvement cannot only be attributed to our accounting of triaxial deformation but also reflects the different BSkG3 predictions for ground state SPL and MOIs for axially symmetric and even spherical nuclei.

\begin{figure}[htbp]
 \centering
\includegraphics[scale=0.4]{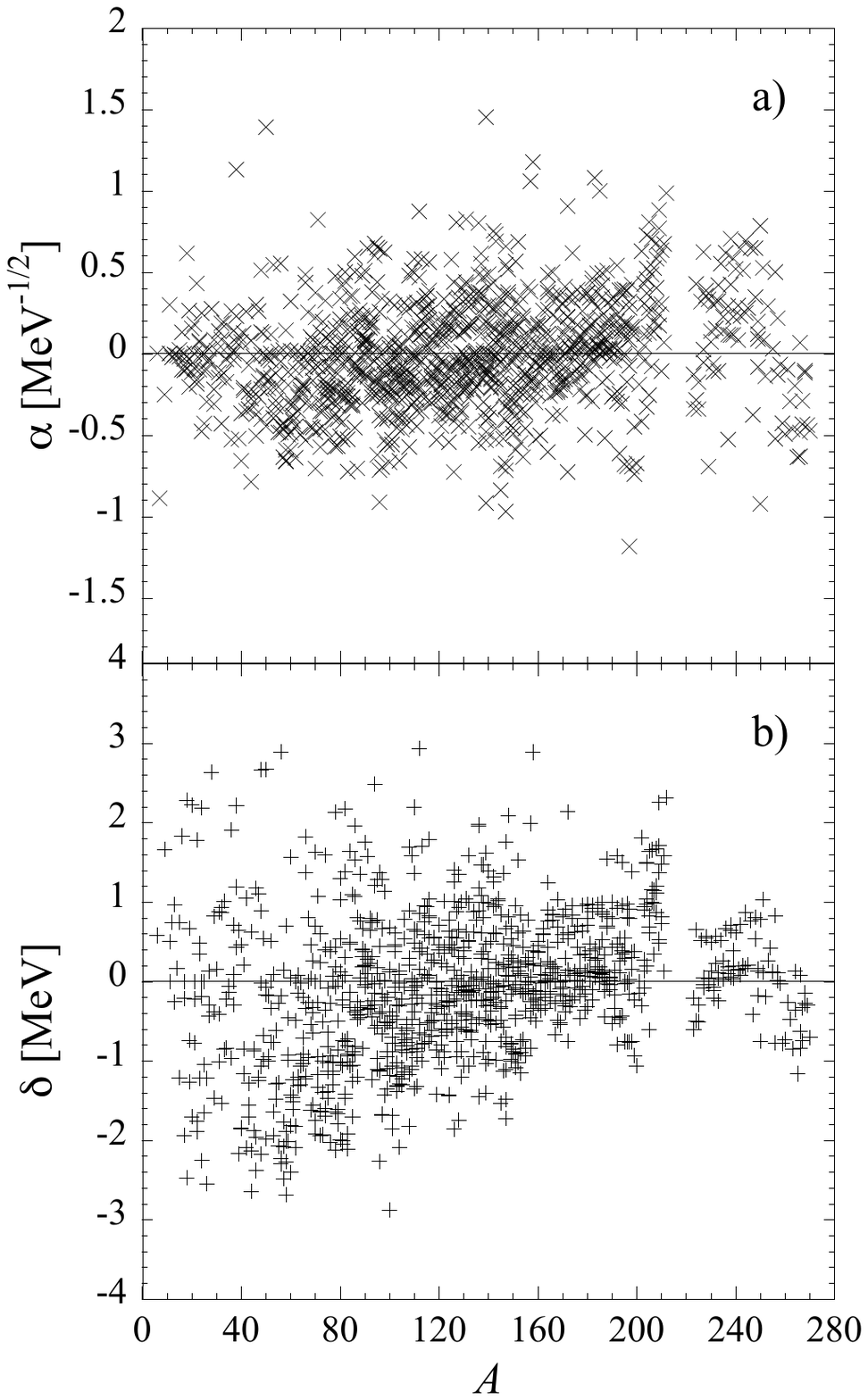}
\caption{Renormalisation parameters $\alpha$ (panel a)and $\delta$  (panel b), as defined in Eq.~\eqref{eq_renorm}, plotted as a function of the atomic mass $A$. See text for more details.}
\label{fig_cptable}
\end{figure}

 For many nuclear physics applications, this level of performance is not sufficient, in particular for nuclear data evaluation or for an
 accurate and reliable estimate of reaction cross sections. Though the HFB plus combinatorial NLD are provided in a
 table format, it is possible to renormalize them on both the experimental level scheme at low energy and the neutron
 resonance spacings at $U=S_n$ in a way similar to what is usually done with analytical formulae. More specifically,
 the renormalized level density can be corrected through the expression
 \begin{equation}
 \rho(U,J,\pi)_{\rm renorm}=e^{\alpha\sqrt{(U-\delta)}} \times \rho(U-\delta,J,\pi)
 \label{eq_renorm}
 \end{equation}
 where the energy shift $\delta$ is essentially extracted from the analysis of the cumulative number of levels 
 and $\alpha$ from the experimental s-wave neutron spacing. With such a renormalization, the experimental low-lying
 states and the $D_{\rm exp}$ values can simultaneously be reproduced at a reasonable level, as discussed in detail in Ref.~\cite{Koning08}. Eq.~(\ref{eq_renorm})
 has been used to fit the 299 nuclides for which both an experimental s-wave spacing ($D_0$) and a discrete level sequence exist. For an additional 830 nuclides,  the experimental discrete level scheme with at least 10 levels is known. For those nuclei,  the $\delta$ shift can be estimated to reproduce at best the low-lying levels, but at the same time, keeping the NLD prediction from the original BSkG3 plus combinatorial predictions, the $\alpha$ parameter needs to be renormalized. 
 The corresponding  $\delta$ and $\alpha$ values are shown in Fig.~\ref{fig_cptable}. 

%------------------------------------------------
%--------------------------------------------------

\begin{figure*}
\begin{center}
\includegraphics[scale=0.5]{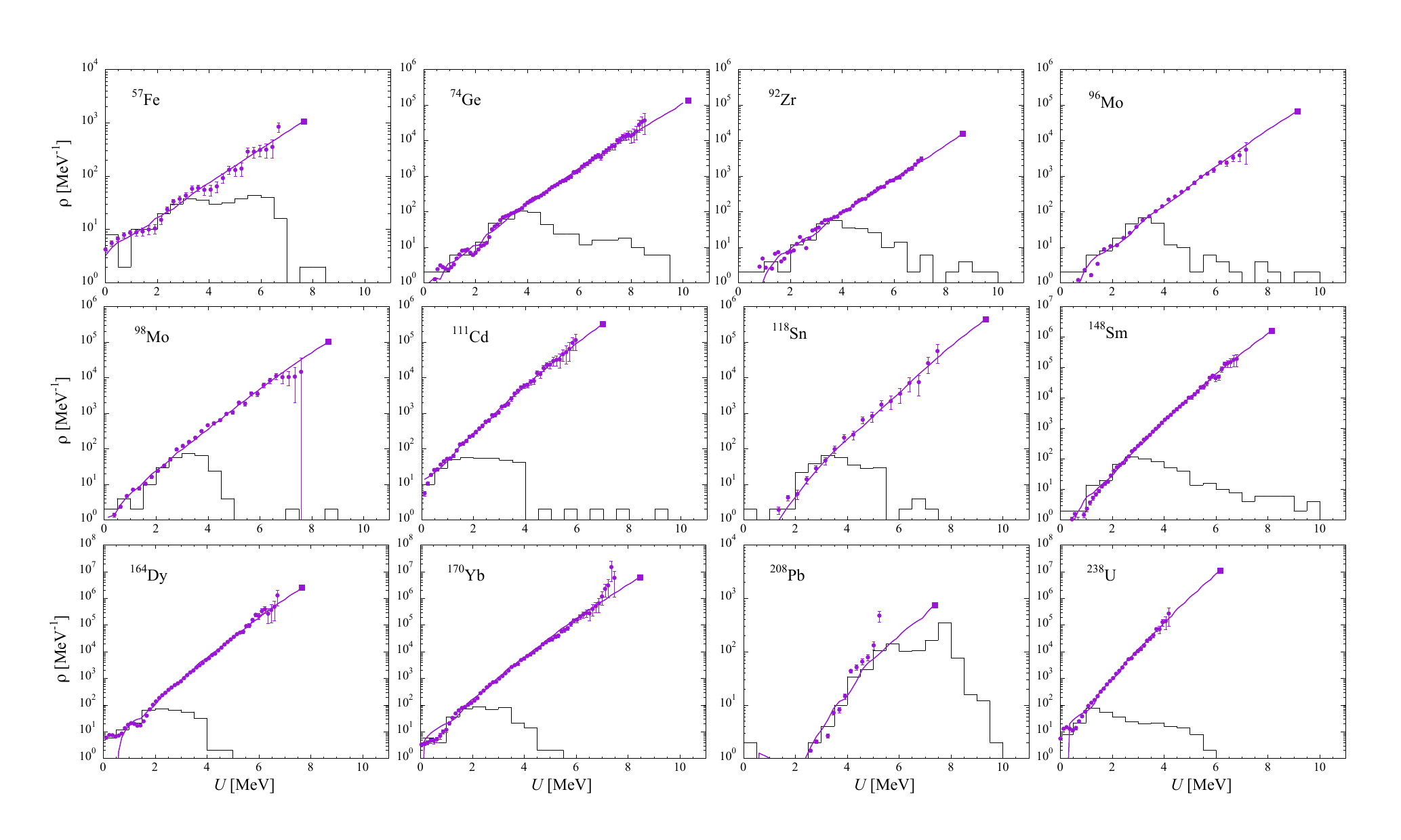}
% \protect
%  \vspace{-1.0cm}
 \caption{Comparison of the BSkG3 plus combinatorial total level densities with those extracted from the Oslo method (dots) \cite{Oslo}. The black line shows the total level density derived from known low-lying levels \cite{Capote09} and black squares the total level density at $U=S_n$ determined from the experimental s-wave spacings \cite{Capote09}. The Oslo data are renormalized as detailed in Ref.~\cite{Goriely22a}. Note that among these nuclides, $^{57}$Fe, $^{74}$Ge, $^{96,98}$Mo, $^{111}$Cd and $^{148}$Sm ground states are predicted to be triaxial with a deformation $15^\circ \le \gamma \le 45^\circ$. }
\label{fig_oslo}
\end{center}
\end{figure*}

 Finally, we compare in Fig.~\ref{fig_oslo} our NLD calculations with the experimental data extracted by the Oslo group
 \cite{Oslo}. The total NLD $\rho_{\rm tot}(U)=\sum_{J,\pi} \rho(U,J,\pi)$
 is compared with our microscopic results in Fig.~\ref{fig_oslo} for 12 nuclides. Since the Oslo data are model-dependent, they have been renormalized following the procedure described in Ref.~\cite{Goriely22a}. The energy dependence of our BSkG3 plus combinatorial model is seen to fairly reproduce the experimental data, even at the lowest energies. The difficult case of $^{208}$Pb is also in good agreement with both Oslo data and the experimental low-lying level scheme.

 \section{Application to the cross section calculation} 
 \label{sect_xs}
 
\begin{figure}[htbp]
\includegraphics[scale=0.30]{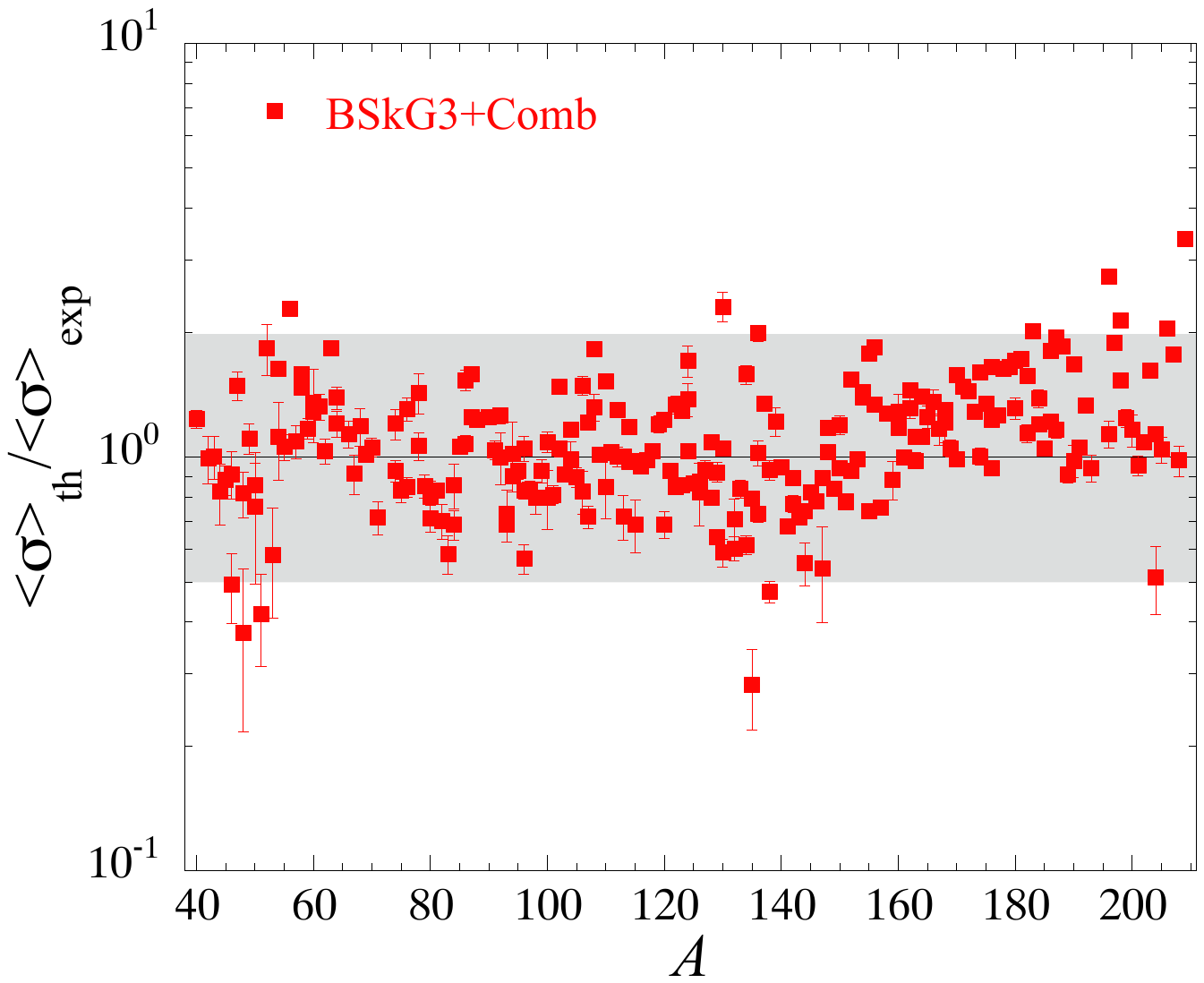}
\caption{Ratio of the theoretical to experimental MACS at  $kT=30$~keV as a function of the atomic mass $A$ for all the 236 nuclei between Ca and Bi for which  an experimental MACS exists  \cite{Dillmann06}. The MACS have been obtained with the present BSkG3 plus combinatorial NLD and the D1M+QRPA photon strength functions \cite{Goriely19}. }
\label{fig_macs}
\end{figure}

 To test the new NLDs, the Maxwellian-averaged neutron capture cross sections (MACS) have been calculated systematically on the basis of the Hauser-Feshbach statistical model  described by the TALYS reaction code \cite{Koning23}. The BSkG3 predictions have been consistently input in the description of all ground-state structure properties. In addition to the present NLDs, the neutron optical potential of Ref.~\cite{Koning03} and the photon strength function from D1M+QRPA+0lim model \cite{Goriely18a} are used. The MACS are compared in Fig.~\ref{fig_macs} with experimental data  \cite{Dillmann06} for 236 nuclei lying between Ca and Bi. An overall deviation $f_{\rm mean}=1.09$ and $f_{rms}=1.43$ is obtained (using rms deviation given by Eq.~(\ref{eq_frms2}). This result can be compared with the one obtained using the BSk14 plus combinatorial NLD \cite{Goriely08b} for which $f_{\rm mean}=1.22$ and $f_{rms}=1.54$.

%------------------------------------------------
%--------------------------------------------------

\begin{figure}[htbp]
\includegraphics[scale=0.3]{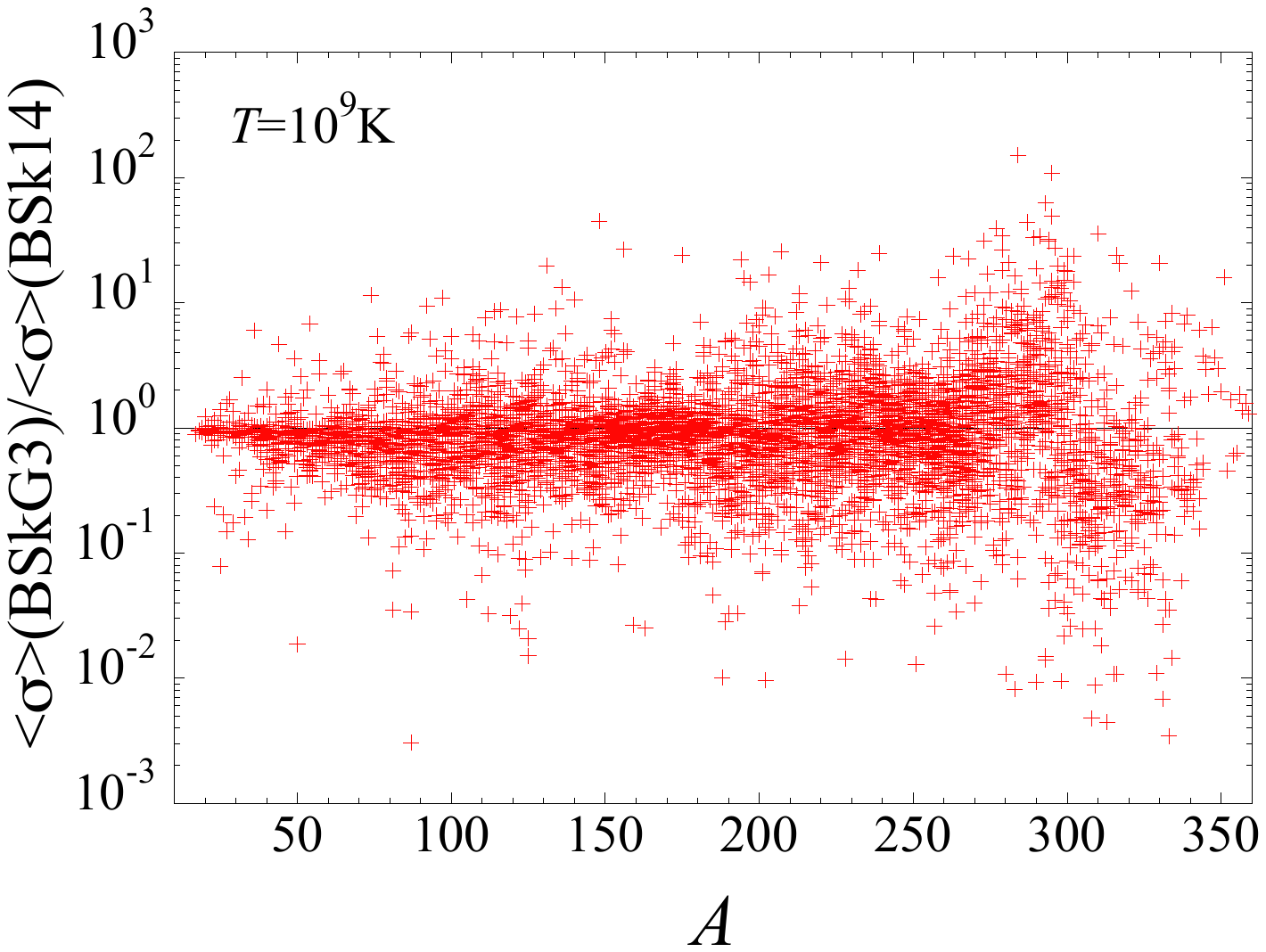}
\caption{Ratio of the MACS at  $T=10^9$~K obtained with the present BSkG3 plus combinatorial NLD to those obtained with the BSk14 plus combinatorial NLD \cite{Goriely18b} as a function of the atomic mass $A$ for all nuclei with $10 \le Z \le 110$ lying between the BSkG3 proton and neutron drip lines.  }
\label{fig_macsa}
\end{figure}

\begin{figure}[htbp]
\includegraphics[scale=0.25]{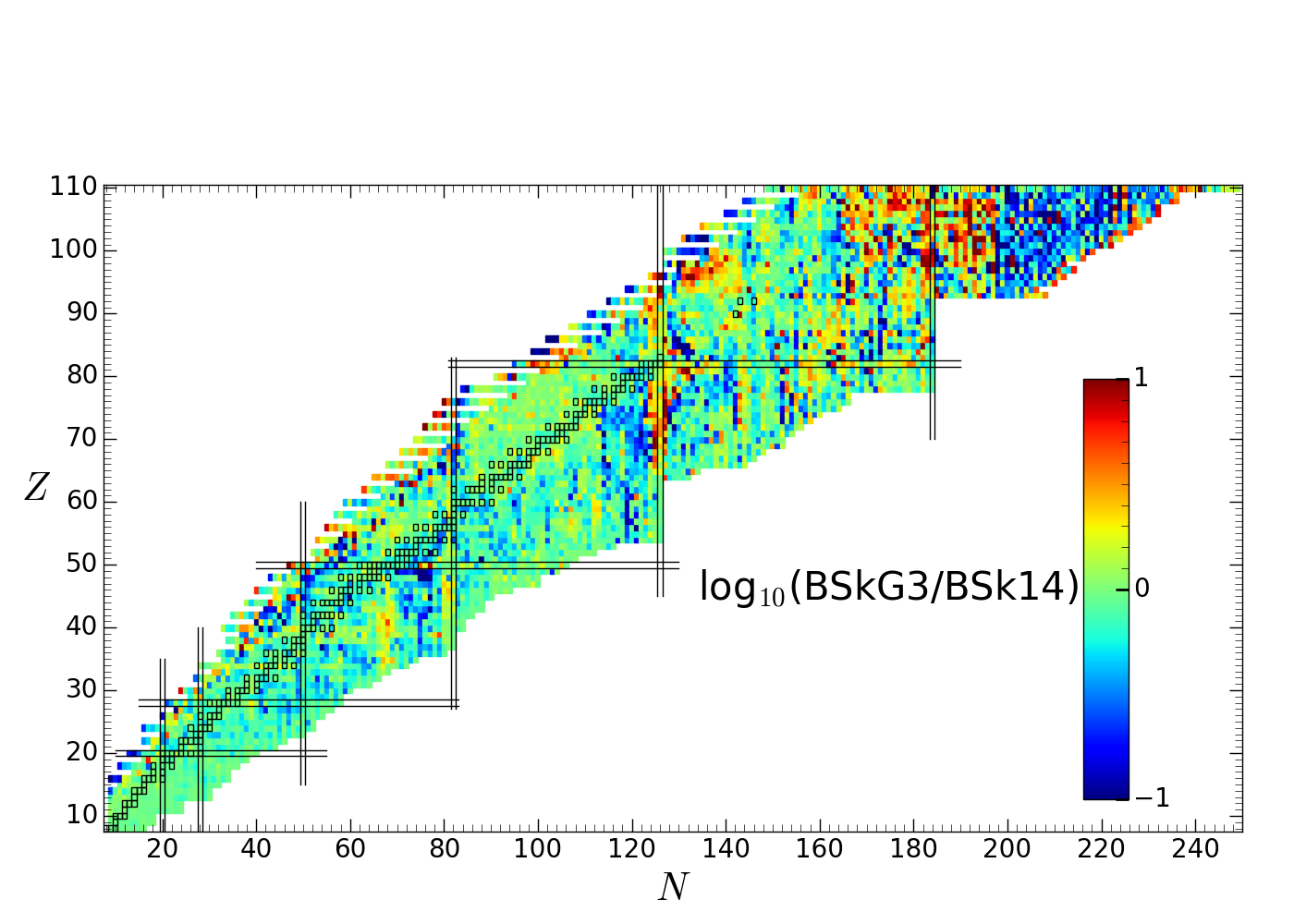}
\caption{Same as Fig.~\ref{fig_macsa} represented in the ($N,Z)$ plane. }
\label{fig_macsnz}
\end{figure}

The neutron MACS have been calculated with the present BSkG3 plus combinatorial NLD for all nuclei with $10 \le Z \le 110$ lying between the BSkG3 proton and neutron drip lines and compared with those obtained with our previous BSk14 plus combinatorial NLD in Figs.\ref{fig_macsa}-\ref{fig_macsnz}. Deviations between both MACS are essentially found within a factor of 10 but can reach a factor 100 up or down. The most affected nuclei are seen to correspond either to exotic n-rich or n-deficient nuclei. In particular, BSkG3 NLDs tend to give larger cross sections around the $N=126$ and 184 magic number but lower cross sections for super-heavy n-rich nuclei. These discrepancies are not only due to the effect of triaxiality but more globally to the respective predictions of the ground state structure properties, such as the deformation, the SPL scheme, as well as the pairing strength. 

%------------------------------------------------
%--------------------------------------------------

%------------------------------------------------
%--------------------------------------------------

\section{NLD at the fission saddle points}
\label{sect_fis}
The BSkG3 plus combinatorial method developed to estimate the NLD at ground-state deformation can also be applied to the calculation of the NLD for the fission saddle points and fission isomers, making use of the corresponding BSkG3 SPL scheme at the corresponding deformation.For actinide nuclei, the inner barrier is often triaxial and the outer barrier octupole deformed (and often triaxial too)~\cite{Ryssens22}; the effects of these two types of deformation on both the fission path and the NLD can be taken care of by the BSkG3 plus combinatorial calculation consistently. In the case of an octupole ground state or saddle point, since parity is not conserved, the NLD is calculated for one parity and assumed to be equal for the other parity (Sect.~\ref{sect_intrinsic}), a factor of two being added to the NLD of both parities for the octupole effect on the collective enhancement \cite{Bjornholm80}. At the saddle points, in contrast to the ground-state configuration, the nucleus is always strongly deformed, so that no damping function is applied, i.e. ${\cal F}=1$ in Eq.~\eqref{eq:fdam}.

Although the NLD is well constrained by the HFB structure properties, the determination of the associated vibrational enhancement remains uncertain. Due to the lack of observables, the
same prescription is used for the saddle points as for the ground state, i.e.,
a total of three phonons have been coupled to the excited configurations of a
maximum of up to four ph pairs. Quadrupole, octupole, and
hexadecapole phonons are included, their energies being assumed identical to
those of the ground state. This prescription leads to a damping of the
NLD vibrational enhancement factor at a relatively low energy (typically 10~MeV).
This damping prescription is known to have a rather significant impact on the first-chance fission cross section at energies above typically 10~MeV \cite{Goriely11c}.

\begin{figure}
    \centering
    \includegraphics[scale=0.43]{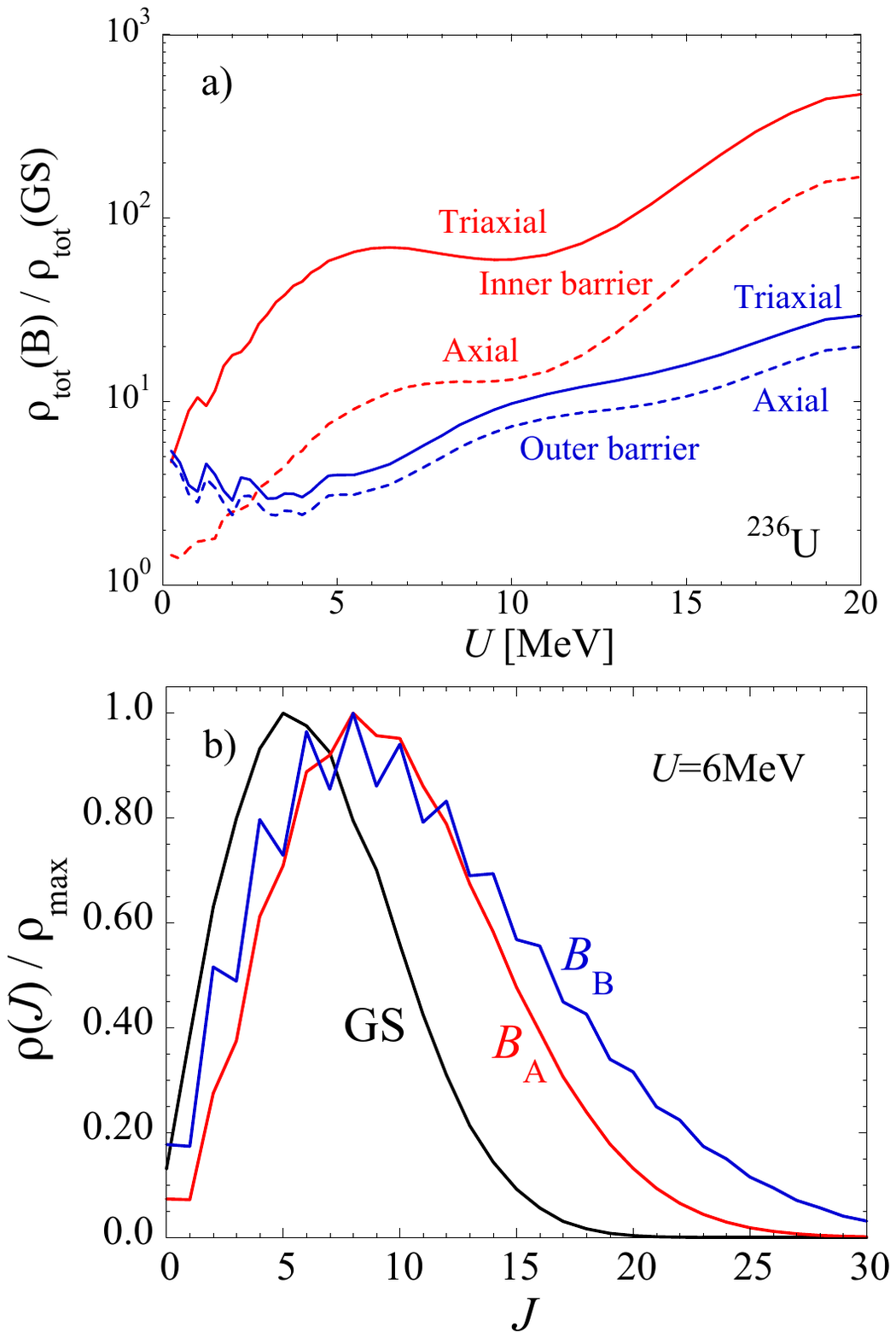}
    \caption{a)  Total NLD at the $^{236}$U inner ($B_A$) and outer ($B_B$) barriers relative to the ground-state (GS) total NLD as a function of the excitation energy $U$. The solid lines correspond to the triaxial configurations and the dashed lines to the axially symmetric saddle points. b) Spin distribution at $U=6$~MeV at the GS, inner and outer saddle points. }
    \label{fig_nld_239U_Bf_vs_GS}
\end{figure}

The total NLDs of $^{236}$U at the first and second barriers, relative to the ground-state total NLD,  are shown in Fig.~\ref{fig_nld_239U_Bf_vs_GS}. The NLD at the outer barriers is seen to be a factor of about 3--5 (or more) larger at low energies than the ground-state NLD; in this case the impact of triaxiality is limited as it increases the NLD by about 30-40\% compared to an axial calculation with identical SPL. In contrast, a factor of about 60 is rapidly reached at the largely triaxial inner barrier; here the triaxial effects are responsible for an increase by a factor of about 7. In addition, the enhancement with respect to the ground-state NLD is also due to the larger deformation of the saddle points; this translates to a low single-particle density near the Fermi energy, increased pairing effects and increased MOIs. In particular, the larger MOI gives rise to a  wider spin distribution, as seen in Fig.~\ref{fig_nld_239U_Bf_vs_GS}b) and also a larger total NLD.
%------------------------------------------------
%--------------------------------------------------

\section{Conclusion}
\label{sect_conc}
The combinatorial method introduced in Ref.~\cite{Goriely08b} has been updated with the latest state-of-the-art large-scale calculation of the ground state properties on the basis of the BSkG3 effective interaction. This HFB calculation has been shown to predict masses, radii and fission barriers with a high accuracy but also to 
be able to break the ground state rotational, axial, reflection, and time-reversal symmetry thanks to its three-dimensional coordinate-space representation.
The  resulting NLD are qualitatively similar to those we obtained assuming an axial symmetry, but offer now a consistent description for triaxial nuclei. 
The impact of triaxiality is shown to decrease the NLD for slightly deformed nuclei due to the smaller SPL density in the triaxial minimum. However, for well-deformed nuclei, the collective enhancement counterbalances the single-particle effect, so that the resulting NLD is larger than when assuming axial symmetry. 

 The final NLD (without renormalization on experimental data) are made available to the scientific community at the website {\it http://www.astro.ulb.ac.be}. 
The tables include  the
 spin- and parity-dependent NLD for more than 8500 nuclei ranging from $Z=8$ to $Z=118$  for a large energy and spin grid  ($U=0$ to 200 MeV and the lowest 50 spins). 

 The NLD have also been implemented in the TALYS reaction code (publicly available at {\it http://nds.iaea.org/talys}) where the normalisation parameters given in
 Sect.~\ref{sect_res}  are also included. 
 As we have shown, when experimental cross sections are available our normalisation procedure globally
improves the agreement with the data. 
%If no experimental data exists, we believe that using our raw NLD is a better alternative than extrapolating, without physical justification,  parameters  such as the level 
% density parameter $a$ which mixes shell, pairing and collective effects.

 Still, some improvements may be required. In particular, the spherical/deformed character for transitional nuclei is not yet under control. In addition, at increasing energies, the shape of the nucleus changes, so that building the excitation configurations on top of the ground state single-particle properties may not be adequate. Such effects will be studied in a near future.

\section*{Acknowledgments}
S.G. and W.R. acknowledge financial support from F.R.S.-FNRS (Belgium). This work was supported by the Fonds de la Recherche Scientifique - FNRS and the Fonds Wetenschappelijk Onderzoek - Vlaanderen (FWO) under the EOS Project No O000422. 
The present research benefited from computational resources made available on the Tier-1 supercomputers Zenobe and Lucia of the Fédération Wallonie-Bruxelles,
infrastructure funded by the Walloon Region under the grant agreement nr 1910247.
Further computational resources have been provided by the clusters Consortium des Équipements de Calcul Intensif (CÉCI), funded by F.R.S.-FNRS under Grant No. 2.5020.11 and by the Walloon Region. 
%W.R. and S.G. gratefully acknowledge support by the F.R.S.-FNRS.
%\end{acknowledgments}
\bibliographystyle{unsrt}
\bibliography{astro}

\clearpage
%%%%%%%%%%%%%%%%%%%%%%%%%%%%%%%%%%%%%%%%%%%%%%%%%%%%%%%%%%%%%%%%%%%%%%%%%%%%%%%%%%%%%%%%%%%%%%%%%%%%%%%%%%%%%%%%%%%%%%%%%%%%%%%%%%%%%%%%%%%%%%%%
\appendix
\section{Shapes and orientation}
\label{sec:orientation}

We rely on the MOCCa code that solves the Skyrme-HFB equations on a three-dimensional coordinate mesh~\cite{ryssens2016}. Such a numerical
representation naturally allows us to consider quite general nuclear configurations; for instance, some of the nuclear ground states predicted
by BSkG3 have finite octupole deformation while all calculations for odd-mass and odd-odd nuclei break time-reversal and account for the influence of
the so-called `time-odd' terms of the Skyrme EDF~\cite{Ryssens2022}. These simulations are nevertheless not symmetry unrestricted: like its predecessor EV8~\cite{ryssens2015}, we impose both $z$-signature $\hat{R}_z$ and $y$-timesimplex $\check{S}^{T}_y$ as self-consistent symmetries to limit the numerical effort required~\cite{dobaczewski2000}. This choice does not
meaningfully impact the generality of our description of nuclear ground states, but has practical consequences for the calculation of NLDs that we describe below.

First, this restricts some of the possible multipole deformations we can study. We define the dimensionless multipole moments of a nuclear configuration:
\begin{align}
\beta_{\ell m } \equiv \frac{4\pi}{3 R_0^{\ell} A^{1+\ell/3}} \int d^3r \, \rho(\vec{r}) r^{\ell} Y_{\ell m}(\theta, \phi) \, .
\end{align}
where $\rho(\vec{r})$ is the nuclear matter density, $R_0 = 1.2 $ fm and $Y_{\ell m}$ is a spherical harmonic.
The self-consistent symmetries we assume put restrictions on several multipole moments; for instance, the
quadrupole moments with $m = \pm 1$ vanish while $\beta_{22}$ and $\beta_{2-2}$ are equal and real. The quadrupole deformation
of a nucleus - even when triaxially deformed - can thus be completely characterized by the numbers $\beta_{20}$ and $\beta_{22}$
or by - perhaps the most widely spread convention - $\beta_{2}$ and $\gamma$, defined as:
\begin{align}
\label{eq:def_beta}
\beta_2 &= \sqrt{\beta^2_{20} + 2 \beta^2_{22}}\, , \\
\gamma &= \text{atan} \left( \sqrt{2} \beta_{22}, \beta_{20} \right)  \, .
\label{eq:def_gamma}
\end{align}
However, there are six equivalent ways to orient a triaxial ellipsoid in a simulation volume with fixed $x$-, $y$- and $z$-axes.
Because of this freedom, not all combinations of $\beta_2$ and $\gamma$ lead to physically distinct shapes; most of the
time, one studies only shapes with $\gamma$ ranging from $0^{\circ}$ to $60^{\circ}$; these extremes correspond to
a prolate shape with rotational symmetry along the $z$-axis and an oblate shape with rotational symmetry along
the $x$-axis. For time-reversal and parity conserving calculations, a single sextant of the $\beta - \gamma$ plane exhausts all physically relevant possibilities.
Strictly speaking, this degeneracy is lifted as soon as we consider configurations with finite angular momentum or left-right asymmetry. We ignore this
subtlety in this paper for reasons of simplicity; in any case, the reorientation effects of angular momentum - so-called `alispin' - are typically less
than 100 keV~\cite{Schunck2010} while essentially all of the left-right antisymmetric ground states predicted by BSkG3 retain axial symmetry~\cite{Grams23}.

Second, exploiting the conservation of $\hat{R}_z$ and $\check{S}^T_y$ for numerical gain forces all single-particle states $\psi$ from which we
build Bogoliubov states to orient their angular momentum along the $z$-axis. More precisely, the single-particle angular momentum expectation values
in $x$- and $y$-directions vanish:%
\begin{align}
  \langle \psi | \hat{J}_x | \psi \rangle  =  \langle \psi | \hat{J}_y | \psi \rangle  = 0\, .
\end{align}
With these conserved symmetries, only the single-particle expectation values of $J_z$ can take nonzero values; $\langle \psi |\hat{J}_z | \psi \rangle$ is
restricted to be a half-integer when the $z$-axis is a rotational symmetry axis, but this single-particle expectation value can take arbitrary real values in general.

Our NLD calculation relies on the introduction of an average quantum number $\bar{K}$ defined with respect to the principal axis with lowest
rotational MOI. Since single-particle angular momenta can only be calculated in one direction, we thus need to select a subset of the
$\beta-\gamma$ plane where we can expect the MOI for rotation around $z$-axis to be the smallest. To do so in a straightforward way, we take Davydov's simple model
for the rotation of a triaxial nucleus~\cite{davydov1958} as guidance: he posits the following $\gamma$-dependency for the rotational MOIs:
\begin{align}
\mathcal{I}^{D}_{x} &= \mathcal{I}_{0} \sin^{2} \left( \gamma - \frac{2\pi}{3}\right) \, , \\
\mathcal{I}^{D}_{y} &= \mathcal{I}_{0} \sin^{2} \left( \gamma - \frac{4\pi}{3}\right )\, , \\
\mathcal{I}^{D}_{z} &= \mathcal{I}_{0} \sin^{2} \left( \gamma \right)\, ,
\label{eq:Iz}
\end{align}
where $\mathcal{I}_0$ is an overall scaling factor. Although the details differ of course from one nucleus to the next, this ansatz for the evolution
of the MOIs with $\gamma$ is qualitatively correct - provided one does not change the total deformation $\beta_2$. The same dependence
is also found in empirical MOIs deduced from Coulomb excitation data~\cite{allmond2017}.

\begin{figure}
\includegraphics[width=.99\columnwidth]{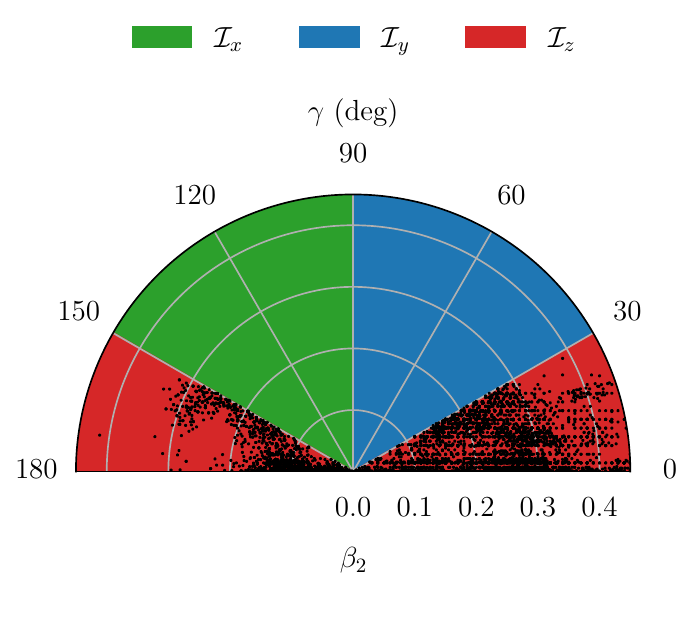}
\caption{
  The smallest rotational MOI ($\mathcal{I}_x, \mathcal{I}_y $ or $\mathcal{I}_z$)
  from the Davydov model as a function of $\gamma$ (see text). The other half of the $(\beta_2, \gamma)$
  plane can be obtained by symmetry. All color-coded regions are equivalent for fully symmetry-unrestricted Skyrme-HFB calculations  but our symmetry assumptions impose constraints on the single-particle states such that only the red region is feasible for our purposes. The black dots mark the location of all nuclei in the BSkG3 mass table; nearly 2000 fall in the $\gamma \in [150^{\circ},180^{\circ}]$ region.}
\label{fig:orientation}
\end{figure}

The Davydov MOI along the $z$-axis is the smallest one for $\gamma \in [0,30^{\circ}] \cup [150^{\circ}, 180^{\circ}]$. For nuclei whose BSkG3
ground state has $\gamma \in [0, 30^{\circ}]$, we can simply perform calculations as in the original publication~\cite{Grams23}. For ground states
with $\gamma \in [30^{\circ}, 60^{\circ}]$ the default orientation of Ref.~\cite{Grams23} is not suitable to NLD calculations; for all of those nuclei
we repeated calculations and ensured that the resulting $\gamma$ fell in the interval $[150^{\circ}, 180^{\circ}]$. For the even-even nuclei whose
ground states are not octupole deformed, these new calculations are exactly equivalent to the original ones of Ref.~\cite{Grams23}; for odd-mass and
odd-odd nuclei as well as even-even nuclei with finite octupole deformation however, our calculations differ from the original BSkG3 mass table by
small rearrangement effects. The final distribution of nuclei in the BSkG3 mass table is illustrated in Fig.~\ref{fig:orientation}.

\section{Root-mean-square deviation indicators}
\label{sect_appendix}

When dealing with the ratio of theory to experiment for a set of data points, the rms deviation is traditionally generalized by the $f_{\rm rms}$ deviation function
\begin{equation}
f_{\rm rms} = \exp  \left[ \frac{1}{N_e} \sum_i^{N_e} \left( \ln r_i \right) ^2 \right] ^{1/2}
\label{eq_frms}
\end{equation}
and the asymmetry as
\begin{equation}
f_{\rm mean} = \exp \left[ \frac{1}{N_e} \sum_i^{N_e} \ln  r_i  \right] ,
\label{eq_erms}
\end{equation}
where $N_e$ is the number of data points and $r_i$ is the ratio between a theoretical to experimental quantity, for example cross sections $\sigma$ or s-wave spacings $D$.
The simplest assumption for $r_i$ is to disregard the experimental uncertainty, often done when the deviation between
model and experiment is large with respect to the experimental uncertainty. Then for a single point we simply have
\begin{equation}
r_i = \frac{\sigma^i_{\rm th}}{\sigma^i_{\rm exp}},
\label{CE}
\end{equation}
which is often referred to as the C/E value.
The resulting $f_{\rm rms}$ value has the advantage over the standard $\chi ^2$ to be a direct measure of the relative deviation. 
A value of $f_{\rm rms}$=1.2 means that for the entire data set the predictions are roughly 20\% off on average from the  central values of the experimental data set.
The asymmetry estimator of Eq. (\ref{eq_erms}) also plays an important role to estimate the goodness of the fit: $f_{\rm rms}$ may have a value significantly different from 1, but as long as $f_{\rm mean}$ is close to 1, we can at least expect that there is no bias 
towards persistent under- or over-estimation of the data by the model.
The disadvantage of Eq. (\ref{CE}) is that the experimental uncertainties are not taken into account.

\begin{figure}[htbp]
\includegraphics[scale=0.35]{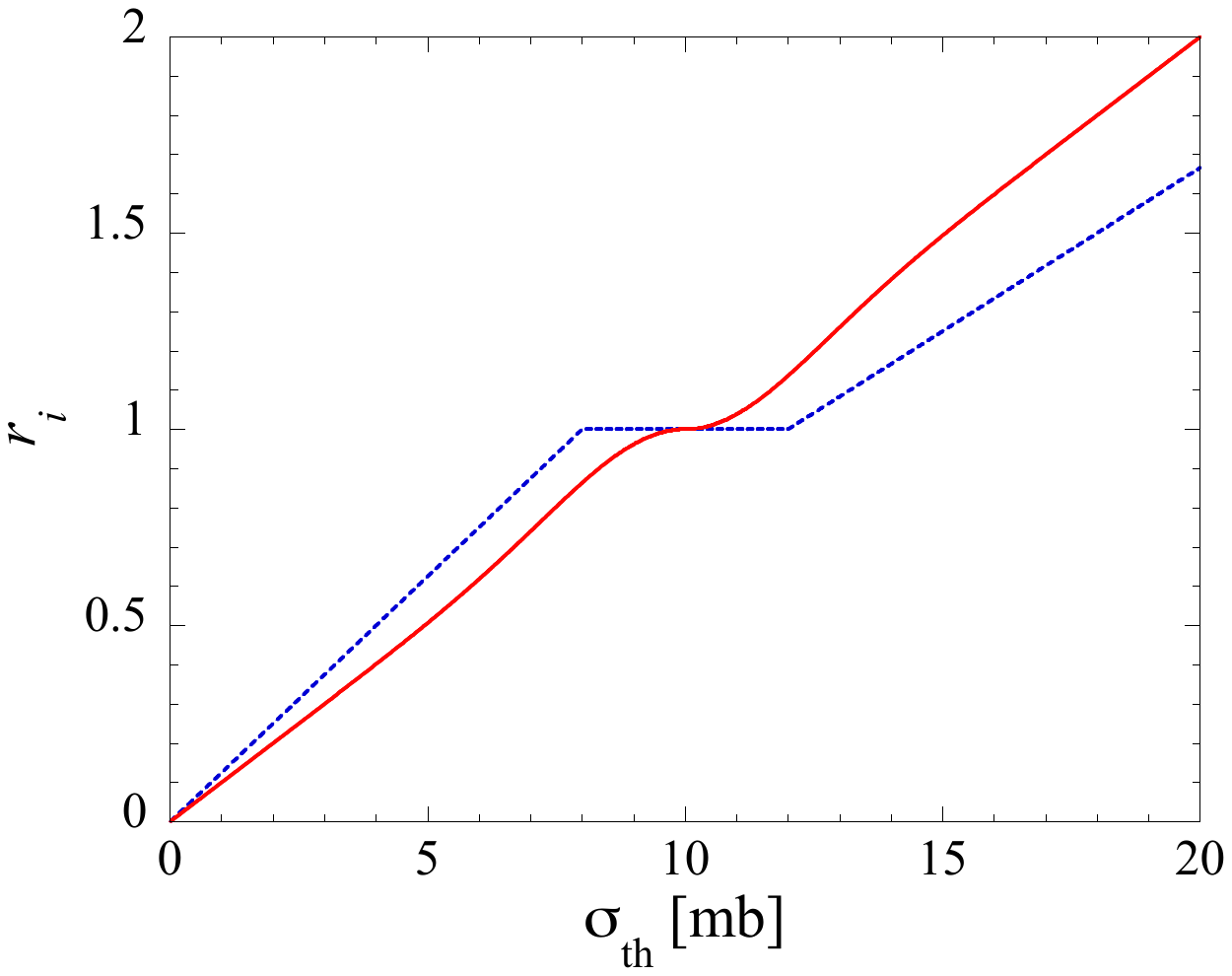}
        \caption{Different $r_i$ estimators for an experimental data point of $\sigma^i_{\rm exp}=10 \pm 2$~mb. The blue dotted line corresponds to Eq.~\eqref{eq_frms1} and the red solid line to Eq.~\eqref{eq_frms2}.}
        \label{fig:CEplot}
\end{figure}

Ideally, one would have an estimator which integrates both the theory-to-experiment deviation from 1 and the experimental uncertainty in one and the same estimator.   To include the impact of experimental uncertainties, one possible way \cite{Goriely18a} is to replace $r_i =\sigma^i_{\rm th} / \sigma^i_{\rm exp}$ by
\begin{eqnarray}
r_i & = & \frac{\sigma^i_{\rm th}}{\sigma^i_{\rm exp}-\delta\sigma^i_{\rm exp}}\ \ \ {\rm if}\ \ \  \sigma^i_{\rm \rm th} < \sigma^i_{\rm exp}-\delta\sigma^i_{\rm exp}, \nonumber\\
    & = & \frac{\sigma^i_{\rm th}}{\sigma^i_{\rm exp}+\delta\sigma^i_{\rm exp}}\ \ \ {\rm if}\ \ \  \sigma^i_{\rm th} > \sigma^i_{\rm exp}+\delta\sigma^i_{\rm exp}, \nonumber\\
    & = & 1 \ \ \ \ \ {\rm otherwise},
\label{eq_frms1}
\end{eqnarray}
which basically states that as soon as the theoretical value is inside the 1-$\sigma$ experimental uncertainty band, 
the goodness-of-fit estimator is 1, and therefore all theoretical values inside $\sigma^i_{\rm exp} \pm \delta\sigma^i_{\rm exp}$ represent an equally good fit.
For practical purposes, this is a powerful recipe: a small experimental uncertainty has a stronger weight than a large uncertainty, 
though not as strong as in the case of the standard $\chi ^2$.
The behaviour of this estimator is depicted in Fig. \ref{fig:CEplot}, where we plot the various deviation estimators for an
experimental data point of $\sigma^i_{\rm exp} = 10 \pm 2$~mb, as a function of the theoretical value  ($\sigma^i_{\rm th}$).

There are two deficiencies of Eq. (\ref{eq_frms1}). First, inside the uncertainty band the Gaussian probability distribution prescribes that the probability is largest at the peak, and should not be flat as in the case of Eq. (\ref{eq_frms1}). Related to that, if we want to use our goodness-of-fit estimators in an optimization procedure of nuclear model parameters to aim for the minimum,
the optimization would be considered successful and stop as soon as the value is inside $\sigma^i_{\rm exp} \pm \delta\sigma^i_{\rm exp}$.

Here, we propose to refine the estimate of  $f_{\rm rms}$ by considering, for a given data point, a normal distribution of width $\delta\sigma^i_{\rm exp}$ around the experimental value rather than a flat distribution as given by Eq. (\ref{eq_frms1}). In this case, the probability density function (pdf) of an experimental data point is given by
\begin{equation}
{\rm pdf}(\sigma^i_{\rm th} ) = \frac{1}{\delta\sigma^i_{\rm exp}\sqrt{2}\pi} \exp \left[ -\frac{1}{2} \left( \frac{\sigma^i_{\rm th} - \sigma^i_{\rm exp}}{\delta\sigma^i_{\rm exp}} \right) ^2 \right] .
\end{equation}
The aim is to reward theoretical values closer to the experimental central value also inside the uncertainty band.
Therefore, we use the cumulative density function  (cdf) as a weight for the theoretical deviation from the experimental point.
If we define
\begin{equation}
x = \frac{\sigma^i_{\rm th}-\sigma^i_{\rm exp}}{\delta\sigma^i_{\rm exp}},
\end{equation}
then
\begin{equation}
{\rm cdf}(x) = \frac{1}{2} \left[  1 + {\rm erf}(\frac{x}{\sqrt 2}) \right] ,
\end{equation}
where {\it erf} is the error function. The {\it cdf} cumulative density function represents the probability obtained by integrating the pdf over a certain region.
When theoretical values are close to the central experimental value, the cdf increases relatively fast, while in the 
tail of the pdf, the cdf slowly reaches convergence.

The deviation of theory versus experiment can be expressed as 
\begin{eqnarray}
r_i & = & 1 - ( \frac{\sigma^i_{\rm th}}{\sigma^i_{\rm exp}} -1){\rm erf}(\frac{x}{\sqrt 2}) \ \ \ {\rm if} \ \ \ \sigma^i_{\rm th} < \sigma^i_{\rm exp}, \nonumber\\
    & = & 1 + ( \frac{\sigma^i_{\rm th}}{\sigma^i_{\rm exp}} -1){\rm erf}(\frac{x}{\sqrt 2}) \ \ \ {\rm if} \ \ \ \sigma^i_{\rm th} > \sigma^i_{\rm exp}, \nonumber\\
    & = & 1  \ \ \ {\rm if} \ \ \ \sigma^i_{\rm th} = \sigma^i_{\rm exp}. 
\label{eq_frms2}
\end{eqnarray}
By using the cdf as a measure for the deviation, we obtain $r_i$ values in between $\sigma^i_{\rm th}/\sigma^i_{\rm exp}$ and the values of Eq. (\ref{eq_frms1}) (see Fig. \ref{fig:CEplot}).
It has the required boundary conditions of $r_i = 1$ when $\sigma^i_{\rm th}/\sigma^i_{\rm exp}=1$ (theory is `perfect' regardless of the experimental 
uncertainty)  and $r_i = \sigma^i_{\rm th}/\sigma^i_{\rm exp}$ at large deviations between theory and experiment (the deviation is so large that the 
experimental uncertainty becomes irrelevant). This last expression is adopted in the present work to quantify the deviation of theory with respect to experiment. 

\centering

\end{document}